\documentclass[preprint,12pt]{elsarticle}

\usepackage{graphicx}
\usepackage{amssymb}
\usepackage{amsmath}
\usepackage{dsfont}
\usepackage{subfigure}
\usepackage[utf8x]{inputenc}
\usepackage{float} 
\usepackage{comment}
\usepackage{color}
 
\usepackage{hyperref}

\usepackage{lineno}

\journal{  }

\begin{document}

\begin{frontmatter}


\title{Soliton Collision in Random Seas}


\author[label1]{Hendrik Fischer}
\ead{he.fischer @tuhh.de}

\author[label1]{Marten Hollm}
\ead{marten.hollm @tuhh.de}

\author[label1]{Leo Dostal}
\ead{dostal @tuhh.de}

\address[label1]{Institute of Mechanics and Ocean Engineering, Hamburg University of Technology, 21073 Hamburg, Germany}


 

\begin{abstract}
Although extreme or freak waves are repeatedly measured in the open ocean, their origin is largely unknown. The interaction of different water waves is seen as one reason for their emergence. One way to consider nonlinear waves in deep water is to look at solutions of the nonlinear Schrödinger equation, which plays an important role in the determination of extreme waves. One specific solution is the soliton solution. Therefore the question arises, how nonlinear waves behave as they interact or collide. 
Using a relaxation pseudo spectral scheme for the computation of solutions of the nonlinear Schrödinger equation, the behavior of colliding solitons is studied. Thereby, different wave amplitudes and angles of collision are considered. In addition to this, the influence of an initial perturbation by random waves is studied, which is generated using a Pierson-Moskowitz spectrum.
\end{abstract}





\end{frontmatter}


\section{Introduction}\label{sec:Introduction}
\label{S:1}

The phenomenon of extreme waves was first observed in connection with ocean waves and has been studied since the seventies of the last century. The amplitude of these waves exceeds the amplitude of the surrounding waves by a factor of more than two \cite{pelinovsky2008extreme}. Furthermore the behavior of extreme waves can be characterized by their sudden, unpredictable appearance and disappearance without any trace. These waves can be accompanied by deep troughs in front and behind the crest of the wave \cite{kharif2003physical}.

Thereby, extreme waves are a real danger for ships, offshore structures and all persons present \cite{pelinovsky2008extreme}. The appearance of an extreme wave can also have fatal consequences for ships as well. According to \cite{kharif2003physical}, 22 super tankers sank between 1969 and 1994 with a total of 525 fatalities. 

Although it has been a well-documented phenomenon, which has been intensively studied in recent years, the causes of such waves are widely unknown. The lack of understanding makes it extremely difficult to predict such waves. In \cite{pelinovsky2008extreme} physical mechanisms such as dispersion, refraction, chaotic behavior, Benjamin-Feir instability and soliton wave interaction are mentioned as possible causes of extreme wave development. More recently, the presence of random wind forcing was investigated \cite{leopaper}. There it was also shown that solitons may persist under wind forcing.

The characteristic steepness of extreme waves suggests that nonlinearities have an important influence on the development of extreme waves.
The same result can be concluded from the satellite observations in the wake of the MaxWave project \cite{pelinovsky2008extreme}. These show that extreme waves occur more often than predicted by the linear wave theory, which was confirmed by experiments at Marintek \cite{onorato2004observation}. This implies the necessity to consider nonlinear wave theory. 

In most problems in offshore engineering, it is sufficient to consider the nonlinear Euler equations instead of the Navier-Stokes equations \cite{dias2008}. But the numerical computation of these equations is still very expensive due to the unknown water surface required as a boundary condition. Therefore, a further problem reduction is of particular interest. Zakharov \cite{zakharov:1968} has shown that weakly nonlinear solutions of the one dimensional Euler equations can be reduced to solutions of the nonlinear Schrödinger equation (NLS), which describe complex wave envelopes. 

One solution of the NLS is the soliton solution. Based on the assumption in \cite{pelinovsky2008extreme} that the collision of solitons could be a possible factor in the development of extreme waves, the collision of solitary waves is a point of research. Thereby, Fedele et. al. \cite{fedele2012special} have found that smooth solitary waves appear to interact elastically, but no results are determined for the non-smooth case of waves which are disturbed by an irregular sea. 

A study on emergence of breather rogue waves in random seas was presented in~\cite{wang2018breather}. Thereby, the Peregrine breather, which is another solution of the NLS, has been considered \cite{Peregrine1983water}. Furthermore, it has been shown that the emergence of the Peregrine breather dynamics can be also attributed to a more general context of higher-order soliton interaction \cite{tikan2017universality}.

In this work soliton wave interactions in irregular sea states are investigated with focus on the effect of these disturbances. In this respect, the creation of a disturbed soliton in a realistic random sea is achieved by the application of the well known model of random sea waves described in \cite{dostal:2011,dostal:2012} in conjunction with the soliton solution. This model is basically built on the superposition of harmonic waves resulting from the linear wave theory. 

The work is structured as follows: In section \ref{sec:Modelling} the nonlinear Schrödinger equation and the soliton solution are introduced, the modeling of the interaction of these solitons is described and the model for the generation of random ocean waves is presented. Building on this, section~\ref{sec:Results} evaluates the collision of solitons in regular seas. These results provide the starting point for the subsequent analysis of the collision of disturbed solitons in random seas. Finally, this work ends with a conclusion in section \ref{sec:Conclusions}.

\section{Nonlinear water waves in an irregular sea}\label{sec:Modelling}
As noted above, the time-consuming numerical computation of the Euler equations motivates a further problem reduction.  It can be shown that weakly nonlinear solutions of these equations can be reduced to a complex wave envelope which satisfies the nonlinear Schr\"odinger equation. Such a reduction can be achieved by means of the method of multiple scales presented for example in \cite{leopaper}. An initial disturbance  finally leads to nonlinear waves in irregular seas.

\subsection{Nonlinear Schrödinger equation}
The NLS is derived using the multiple scale method, taking into account terms up to the order $\mathcal{O}(\varepsilon^3)$ with the wave steepness $\varepsilon \ll 1$. The derivation is shown in \cite{leopaper} in detail and leads to the equation for the case of deep water\\
\begin{equation}\label{eq:NLS}
\textup{i}\psi_\tau = \alpha \psi_{\xi\xi} + \beta \left| \psi \right|^2\psi,
\end{equation}
where $\alpha=\frac{\omega_0}{8k^2}$ and $\beta=\frac{1}{2}\omega_0 k^2$. Additionally, $\psi(\xi,\tau) \in \mathbb{C}$ describes the wave envelope, $\xi = \varepsilon(x-c_g t)$ is the scaled spatial coordinate including the deep water group velocity $c_g = \frac{\omega_0}{2k}$ and $\tau = \varepsilon^2 t$ is the scaled time. Furthermore, $k$ is the wave number and $\omega_0$ is the frequency of the carrier wave. 

According to \cite{witt2019inducing}, the evaluation of the free water surface follows in first order from the NLS by
\begin{equation}\label{eq:evaluation_1}
\eta(x,t) = \text{Re}\left[\psi(x,t)\exp(\textup{i}(kx - \omega_0 t))\right]
\end{equation}
and in second order by
\begin{equation}\label{eq:evaluation_2}
\eta(x,t) = \text{Re}\left[\psi(x,t)\exp(\textup{i}(kx - \omega_0 t)) + \frac{1}{2}k \left[ \psi(x,t)\right]^2 \exp (2\textup{i}(kx-\omega_0 t))\right].
\end{equation}
Additionally, the wave period $T$ and the wavelength $\lambda$ are calculated by
\begin{equation}\label{eq:wavelength_period}
T = \frac{2\pi}{\omega_0}, ~~~ \lambda = \frac{g}{2\pi}T^2,
\end{equation}
where $g$ is acceleration due to gravity. 

For all further examinations, $\omega_0$ and $k$ are chosen as $\omega_0=1 \,\textup{rad/s}$ and $k=1/g$. Therefore, the coefficients $\alpha$ and $\beta$ are also determined.

\subsection{Soliton solution}
With respect to the soliton interaction, it is insufficient to consider a stationary solution of the NLS. A solution extended by a motion must be taken into account. According to \cite{wu2010nonlinear, bao2013numerical} this soliton solution can be described by 
\begin{equation}\label{eq:solution_move}
\psi(\xi,\tau) = a_0 \textrm{sech}\left[a_0 \sqrt{\frac{\beta}{2\alpha}}(\xi-\xi_0-v \tau)\right] \exp(\textup{i}(c\xi-w\tau)),
\end{equation}
\[
c = -\frac{v}{2\alpha}, \: w = -\alpha c^2 + \frac{1}{2} \beta a_0^2
\]
with the amplitude $a_0$, spatial shift $\xi_0$ and velocity $v$ of the soliton. The parameters $\alpha$ and $\beta$ are set by the NLS \eqref{eq:NLS} itself. \textcolor{black}{Thereby, $v$ is the velocity of the soliton in the $(\xi,\tau)$ coordinate system, which can be transformed into the $(x,t)$ coordinate system by $v_{x,t}=c_g+\varepsilon v$. }

\begin{figure}[H]
	\centering
	\subfigure{\includegraphics[width=0.54\textwidth]{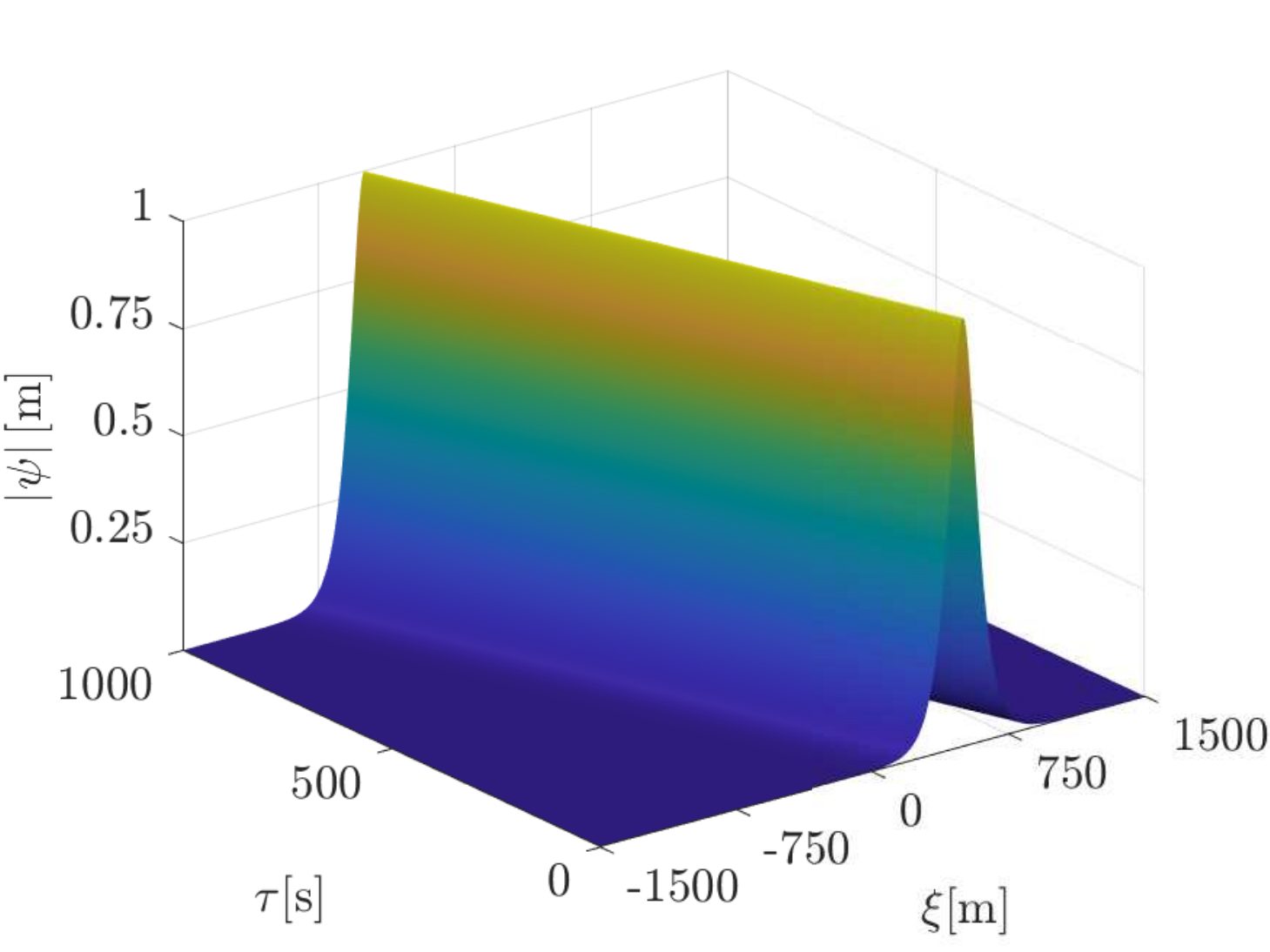}}
	\subfigure{\includegraphics[width=0.45\textwidth]{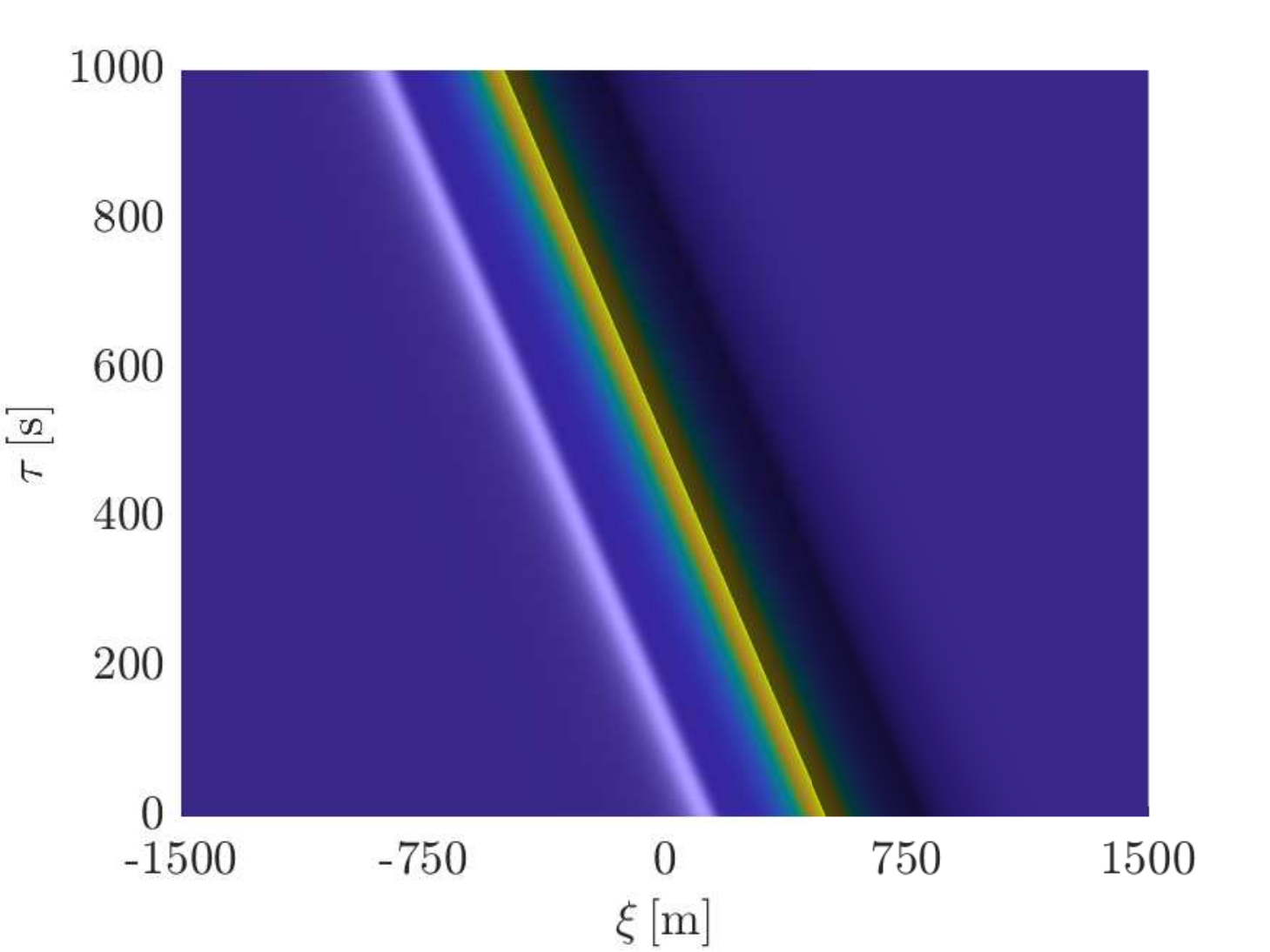}}
	\caption{Analytical soliton solution with amplitude $a_0 = 1 \,\textup{m}$, spatial shift $\xi_0 = 500 \,\textup{m}$ and velocity $v = -1 \,\textup{m/s}$.}
	\label{fig:exact_sol}
\end{figure}

The soliton solution in Fig. \ref{fig:exact_sol} is shown from two different perspectives and is characterized by the parameters $a_0 = 1 \,\textup{m}$, $\xi_0 = 500 \,\textup{m}$ and $v = -1 \,\textup{m/s}$.

\subsection{Description of random ocean waves}\label{sec:RandomOceanWaves}
A well-known model of random sea waves is given by the superposition of harmonic waves with wave numbers $\kappa(\omega)$ and wave frequencies $\omega$ corresponding to a one-sided spectral density $S(\omega)$, cf. \cite{dostal:2011,dostal:2012}. Common sea spectral densities $S(\omega)$ are hereby the JONSWAP spectrum for shallow-water waves and the Pierson-Moskowitz spectrum for deep-water waves. 
In the one dimensional case, beside the amplitude of the waves, the direction of propagation $\gamma$ has to be taken into account too. In order to obtain an initially irregular or random wave surface, a random phase shift $\varepsilon(\omega)$ is added, which is uniformly distributed in $[0,2\pi)$. With this, an irregular wave surface can be written as \\ 
\begin{equation}\label{eq:random_wave_surface_1d}
Z(x,t)=\int_0^{\infty} \int_{-\pi}^{\pi} \cos\left(\omega t -\kappa (\omega) (x \cos(\gamma) )+\varepsilon(\omega)\right) \sqrt{2S(\omega) D(\gamma) \textup{d}\gamma \textup{d}\omega}.
\end{equation}
Here, the integral is not a Riemann integral but a summation rule over the frequencies $\omega$ and the directions of propagation $\gamma$. The propagation function $D(\gamma)$ leads to a scattering of the directions of propagation of the single harmonic waves. The function $D(\gamma)$ has to be normalized in the domain $[-\pi,\pi]$, i. e. 
\begin{equation}\label{eq:norm_prop_function}
\int_{-\pi}^{\pi} D(\gamma)\textup{d}\gamma=1.
\end{equation}
A propagation function $D(\gamma)$, which is often used, is
\begin{equation}\label{eq:example_prop_function}
D(\gamma) \textup{d}\gamma =\frac{2}{\gamma_R} \cos^2 \left( \frac{\pi}{\gamma_R} (\gamma-\gamma_0) \right) ~~ \textup{with } \vert \gamma-\gamma_0 \vert \leq \frac{\gamma_R}{2}.  
\end{equation}
Here, $\gamma_0$ is the main propagation direction and $\gamma_R$ has to be chosen such that Eq. \eqref{eq:norm_prop_function} holds. Under the additional assumption of a long-crested sea state, this can be further reduced to
\begin{equation}\label{eq:random_wave_surface_1d_long-crested}
Z(x,t)=\int_0^{\infty} \cos\left(\omega t -\kappa (\omega) x \cos(\gamma_0) +\varepsilon(\omega)\right) \sqrt{2S(\omega)   \textup{d}\omega},
\end{equation}
since all superpositioned harmonic waves have the same direction of propagation $\gamma_0$. 

In the following calculations the Pierson-Moskowitz spectrum is used which can be described by the significant wave height $H_s$ and the modal frequency $\omega_m$ by
\begin{equation}\label{eq:JONSWAP}
S_{\textup{J}}(\omega) = 0.3125 H_s^2 \frac{\omega_m^4}{\omega^5} \exp\left\{-1.25 \left(\frac{\omega_m}{\omega}\right)^4\right\}.
\end{equation}
\textcolor{black}{If not explicitly mentioned,} in all further investigations the significant wave height $H_s=0.2 \,\textup{m}$, modal frequency ${\omega_m=0.25 \,\textup{rad/s}}$ and direction of propagation $\gamma_0 = 0$ are selected. 

\begin{figure}[H]
	\centering
	\subfigure{\includegraphics[width=0.49\textwidth]{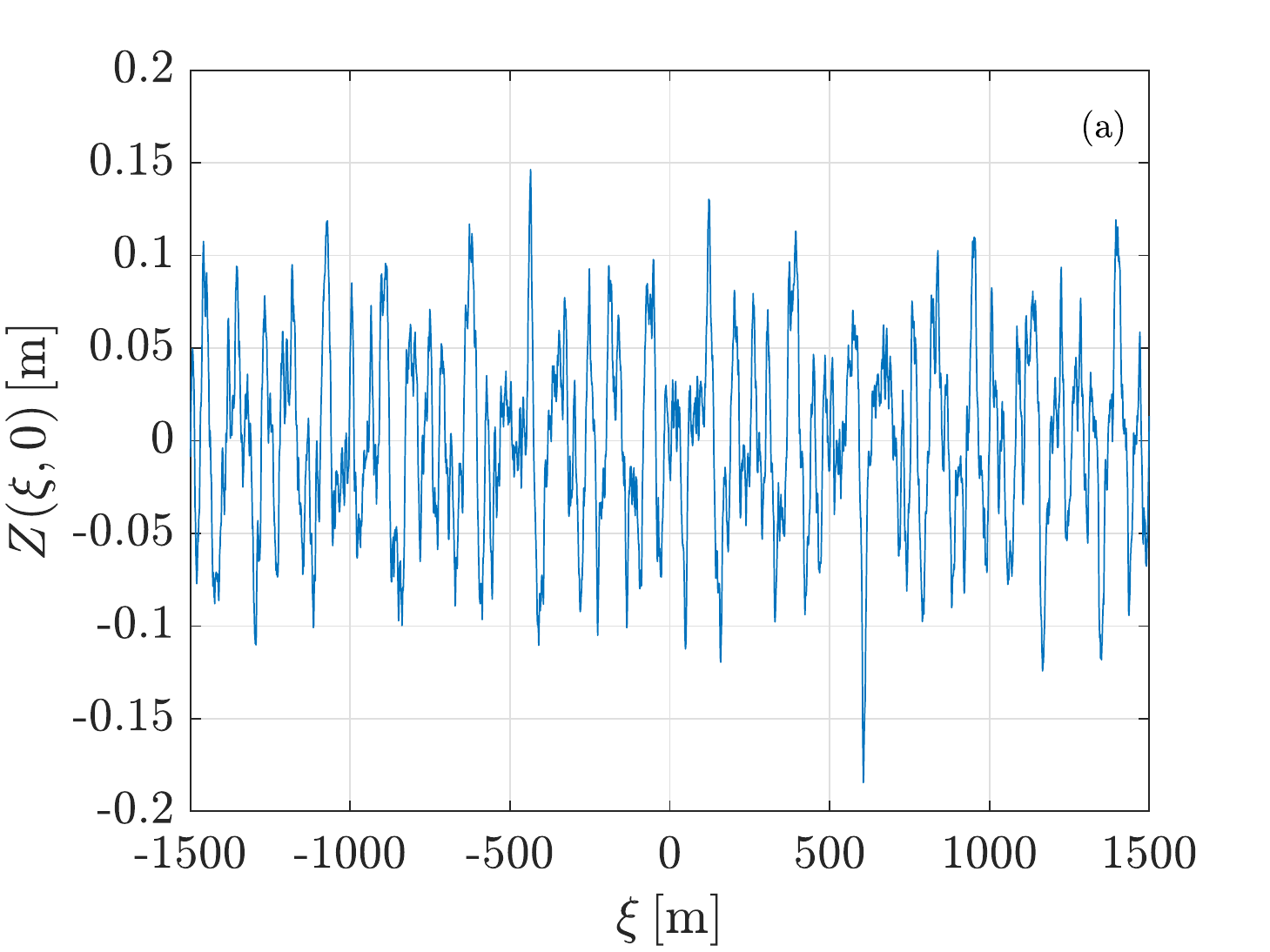}}
	\subfigure{\includegraphics[width=0.49\textwidth]{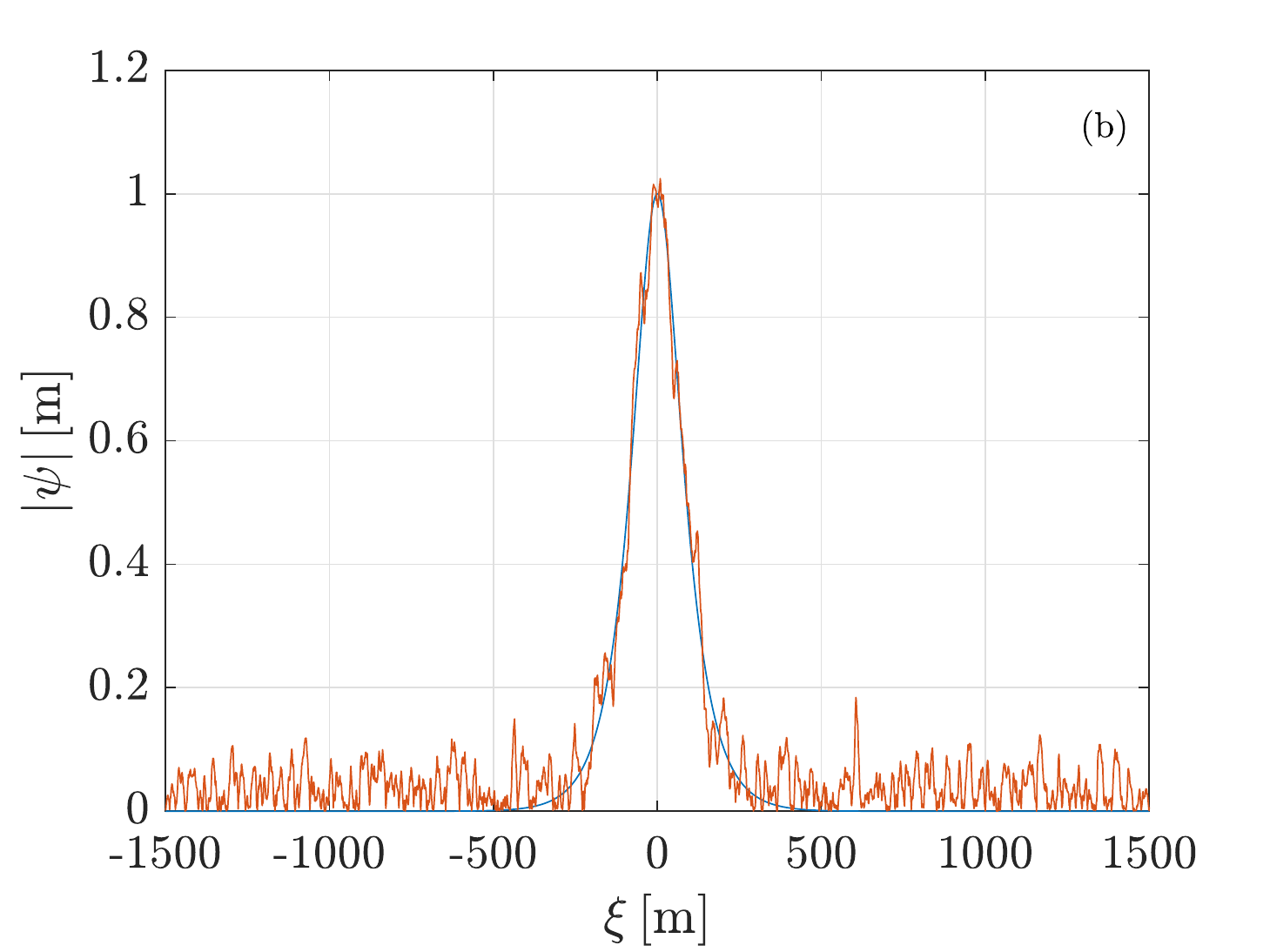}}
	\caption{(a) Generated random ocean sea and (b) the corresponding perturbed soliton solution for $\tau=0$.}
	\label{fig:perturbed_IC}
\end{figure}

Using this model, the initial condition of the numerical simulation, described by the analytical solution \eqref{eq:solution_move}, is adjusted to simulate a realistic irregular sea state. Let $\eta_{\textup{d}}(\xi,0)$ be the \textcolor{black}{initial disturbed free sea surface and $\eta(\xi,0)$ the sea surface obtained from the NLS eq. \eqref{eq:evaluation_1} or \eqref{eq:evaluation_2}, respectively. \textcolor{black}{Here, we use the scaled space-coordinate $\xi$}. In order to achieve 
	\begin{equation}\label{eq:eta_irregular}
	\eta_{\textup{d}}(\xi,0)=\eta(\xi,0)+Z(\xi,0),
	\end{equation}
	we use the following disturbance for $\psi(\xi,0)$:
	\begin{equation}\label{eq:irregular_psi}
	\widetilde{\psi}(\xi,0) = \left( 1+ \frac{Z(\xi,0)}{\vert\psi(\xi,0)\vert} \right) \psi(\xi,0).
	\end{equation}
	By doing so, the amplitude of the disturbed wave elevation can be achieved by
	\begin{equation}\label{eq:irregular_psi_abs}
	\vert\widetilde{\psi}(\xi,0)\vert = \left| \vert\psi(\xi,0)\vert+Z(\xi,0) \right|,
	\end{equation}
	i. e. by adding the amplitude of the undisturbed sea surface and the disturbance $Z(\xi,0)$} and taking the absolute value. Figure \ref{fig:perturbed_IC}a shows such a random sea state calculated by means of Eq. \eqref{eq:random_wave_surface_1d_long-crested} and Fig.~\ref{fig:perturbed_IC}b the application to the soliton solution using Eq. \eqref{eq:irregular_psi}.
%

%
\subsection{Soliton interaction}
For the study of soliton interaction, the initial conditions are selected according to Eq. \eqref{eq:solution_move}. But this equation only describes one soliton. Consequently, a way must be found to combine solitons. For this, the general approach described in \cite{wu2010nonlinear, geiser2019comparison} results in the required superposition of single solitons. Thus, the initial condition for $N$ combined waves is given by
\begin{equation}\label{eq:initial_cond_super}
\psi_0(\xi) = \sum_{i = 1}^N \psi_i(\xi,0),
\end{equation}
whereby $\psi_i(\xi,0)$ is determined by Eq. \eqref{eq:solution_move}. Furthermore, each wave $\psi_i(\xi,0)$ is characterized by its own velocity $v^i$, shift in space $\xi_0^i$ and amplitude $a_0^i$. 

It is important to ensure that the solitons do not already interact with each other in the initial condition in order to guarantee that the entire interaction behavior takes place in the simulation period. This can be done by a suitable choice of the spatial shift $\xi_0^i$  for each wave. 

\begin{figure}[H]
	\centering
	\includegraphics[width=0.75\textwidth]{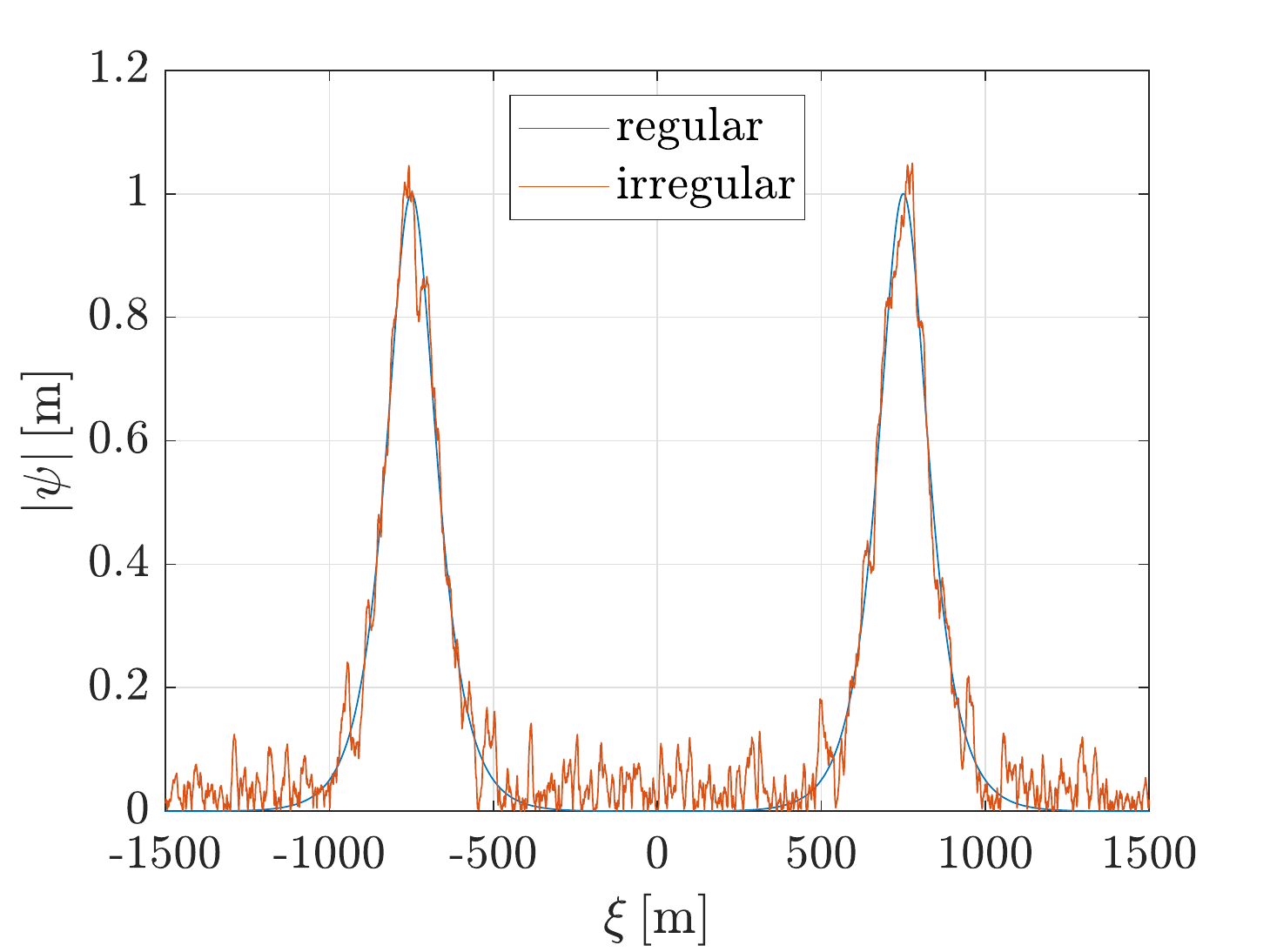}
	\caption{Combined initial condition for two solitary waves for the regular and irregular sea state.}
	\label{fig:perturbed_IC_2sol}
\end{figure}

In order to obtain an irregular or random sea state scenario, the initial condition can be modified using Eq. \eqref{eq:irregular_psi}. This leads to the initial condition
%
\begin{equation}\label{eq:initial_condition_irregular}
\widetilde{\psi_0}(\xi) = \left( 1+ \frac{Z(\xi,0)}{\vert \psi_0(\xi)\vert} \right)\psi_0(\xi)
\end{equation}
which is used for the following calculations. Figure \ref{fig:perturbed_IC_2sol} shows such an initial condition for the regular and irregular case calculated using Eq. \eqref{eq:initial_cond_super} and Eq.~\eqref{eq:initial_condition_irregular}, respectively. 
\section{Results}\label{sec:Results}
Based on the theory presented in section~\ref{sec:Modelling}, this section evaluates the simulations of the soliton interaction. The following simulation results are generated numerically by a relaxation pseudo spectral scheme described in \cite{leopaper}. The main part of this analysis evaluates the maximum of the wave envelope amplitude $\vert \psi \vert$ during the interaction, i.e. 
\begin{equation}
\textup{M}=\max\limits_{\xi, \tau} \vert \psi(\xi,\tau) \vert.
\end{equation}

In the following simulations different scenarios are considered, which differ in wave amplitude and velocity. With regard to the velocities, the magnitude and the direction are of interest. The focus is primary on the interaction of two solitons, since according to \cite{akhmediev2009extreme} the simultaneous collision of several water waves can be assumed as a very unlikely event.

In order to be able to make a general statement about the interaction behavior in irregular seas, the regular case is considered first, i. e. the case without any disturbances. Based on the corresponding results, the scenario with the greatest potential in terms of amplitude development will be analyzed in the irregular case. Due to the stochasticity in the calculation of random ocean waves in subsection~{\ref{sec:RandomOceanWaves}}, an empirical evaluation is of particular relevance for the irregular sea state.

\subsection{Interaction of regular solitons }\label{sec:interaction_regular}
In order to analyze various scenarios, the initial conditions from Eq. \eqref{eq:initial_cond_super} are adjusted accordingly. In this context the choice of the parameters $v^i$, $\xi_0^i$ and $a_0^i$, which specify the velocities, the shifts in space and the amplitudes of the particular waves, is of significant importance.

\begin{figure}[H]
	\centering
	\subfigure{\includegraphics[width=0.54\textwidth]{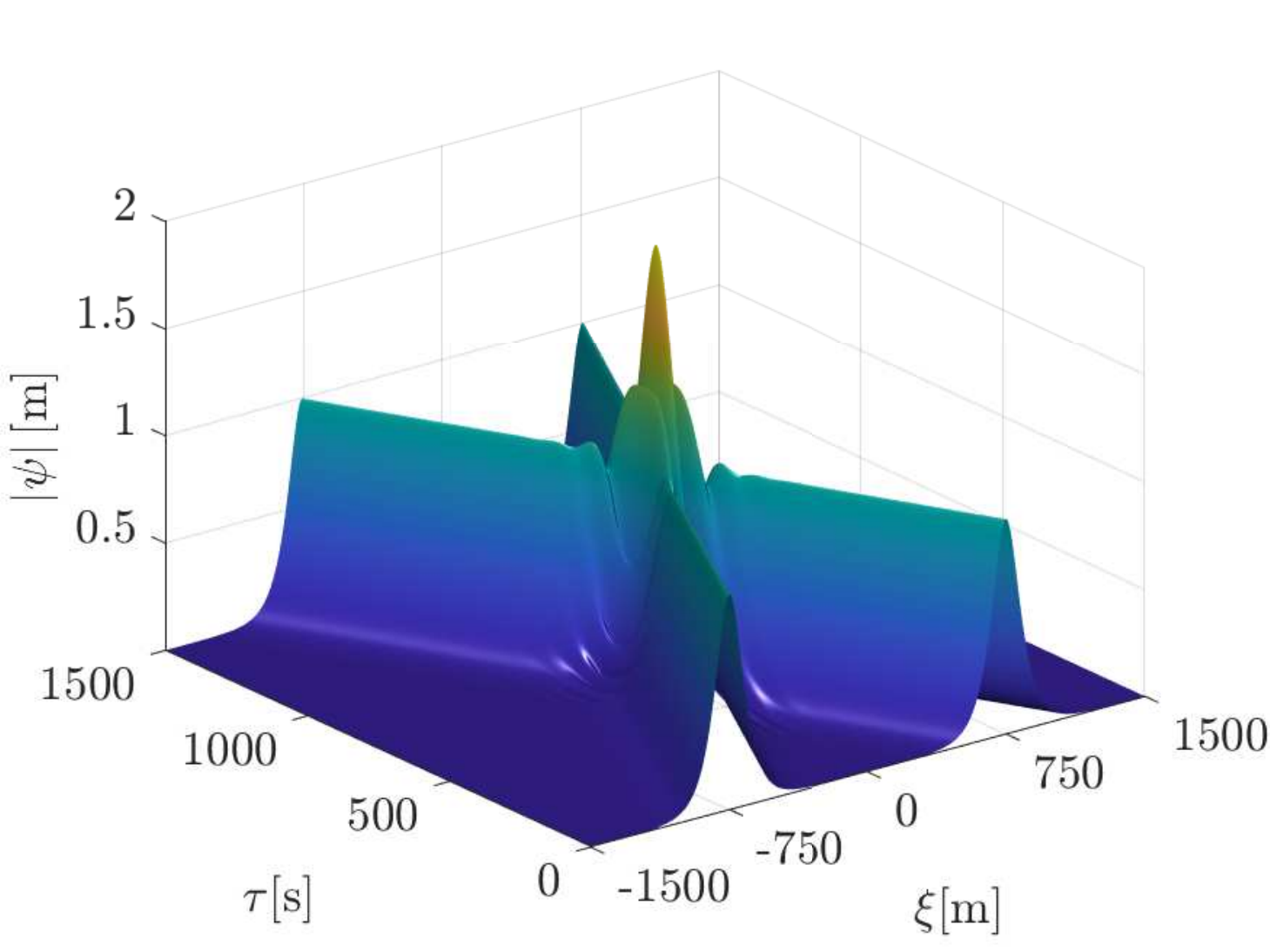}}
	\subfigure{\includegraphics[width=0.45\textwidth]{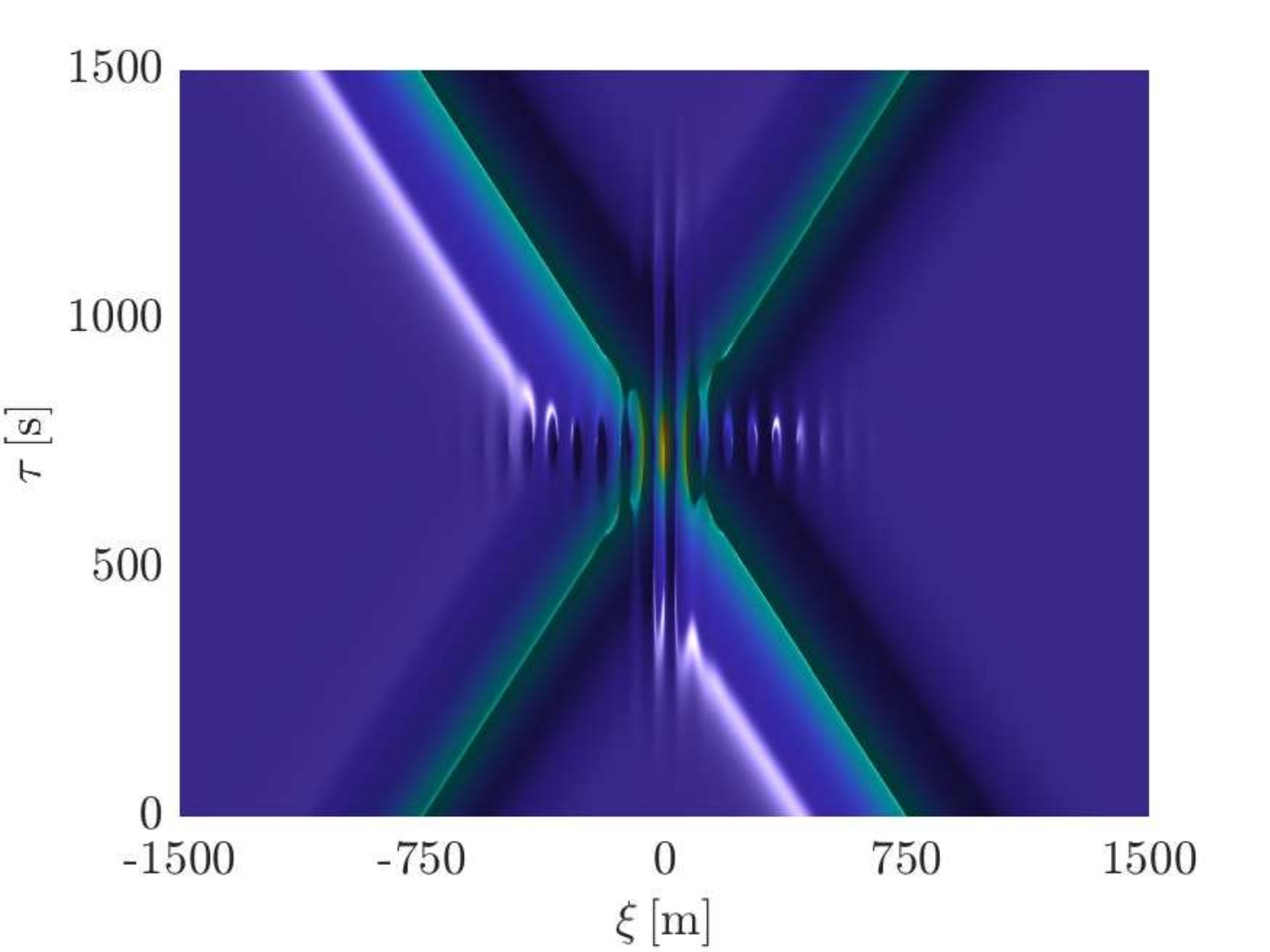}}
	\caption{Interaction behavior of solitons with amplitudes $a_0^1=a_0^2=1 \,\textup{m}$ and velocities $v^1=-v^2= -1 \,\textup{m/s}$ with resulting ${\text{M} =2.0 \,\textup{m}}$.}
	\label{fig:collision_normal}
\end{figure}

\begin{figure}[H]
	\centering
	\subfigure{\includegraphics[width=0.54\textwidth]{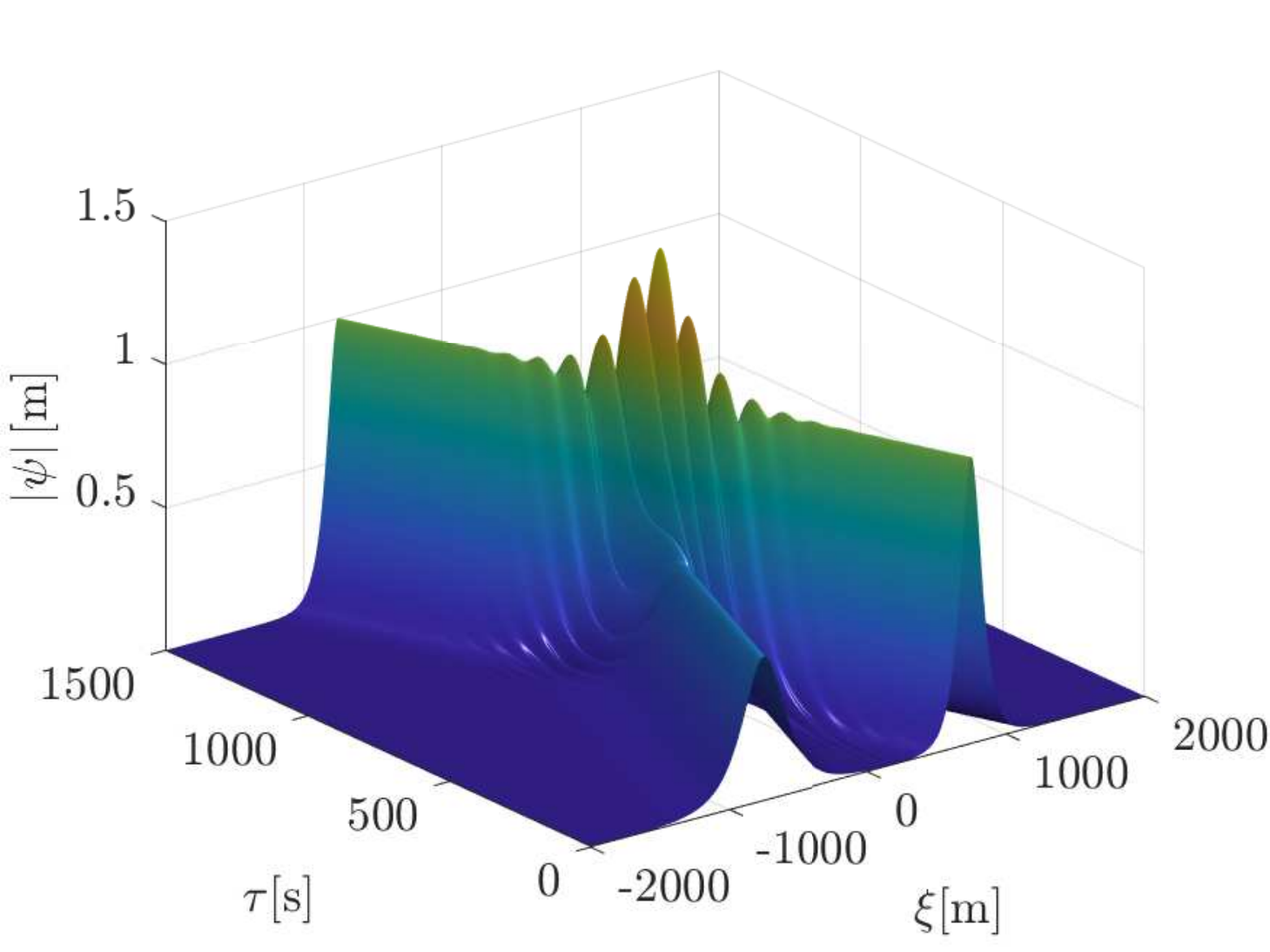}}
	\subfigure{\includegraphics[width=0.45\textwidth]{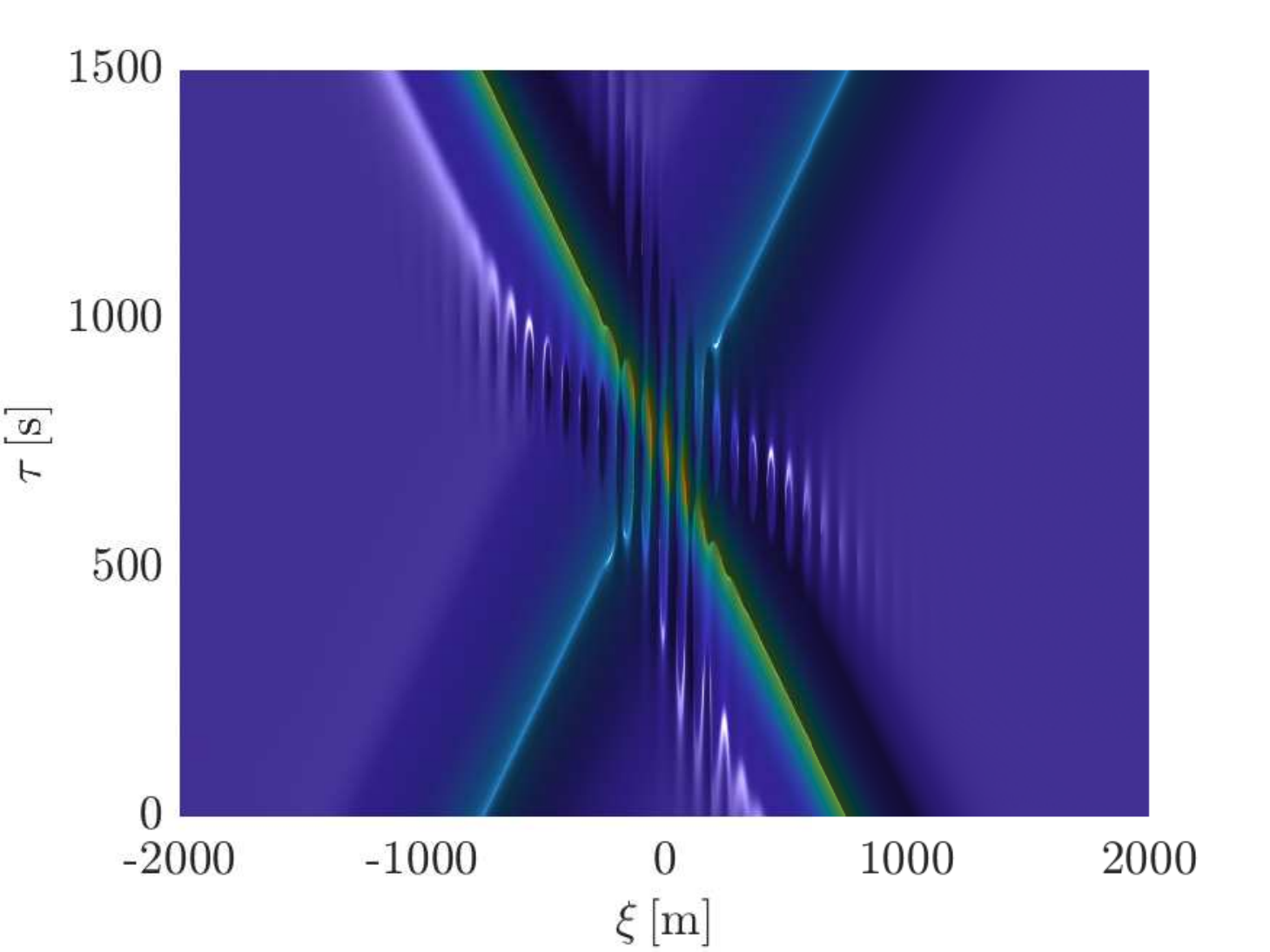}}
	\caption{Interaction behavior of solitons with amplitudes $a_0^1=2a_0^2=1 \,\textup{m}$ and velocities $v^1=-v^2= -1 \,\textup{m/s}$ with resulting ${\text{M} =1.4915 \,\textup{m}}$.}
	\label{fig:collision_normal_ampli_diff}
\end{figure}

\begin{figure}[H]
	\centering
	\subfigure{\includegraphics[width=0.54\textwidth]{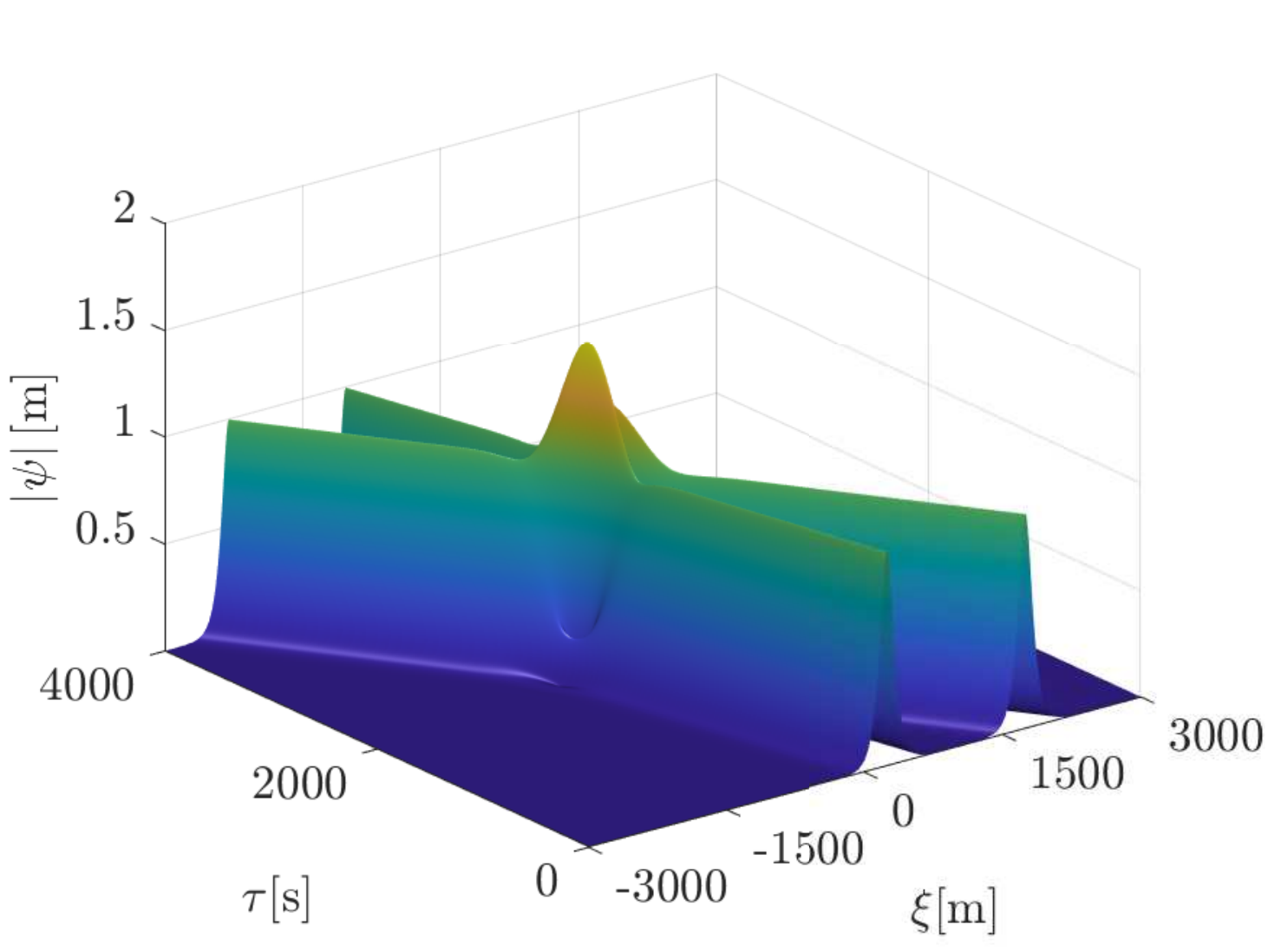}}
	\subfigure{\includegraphics[width=0.45\textwidth]{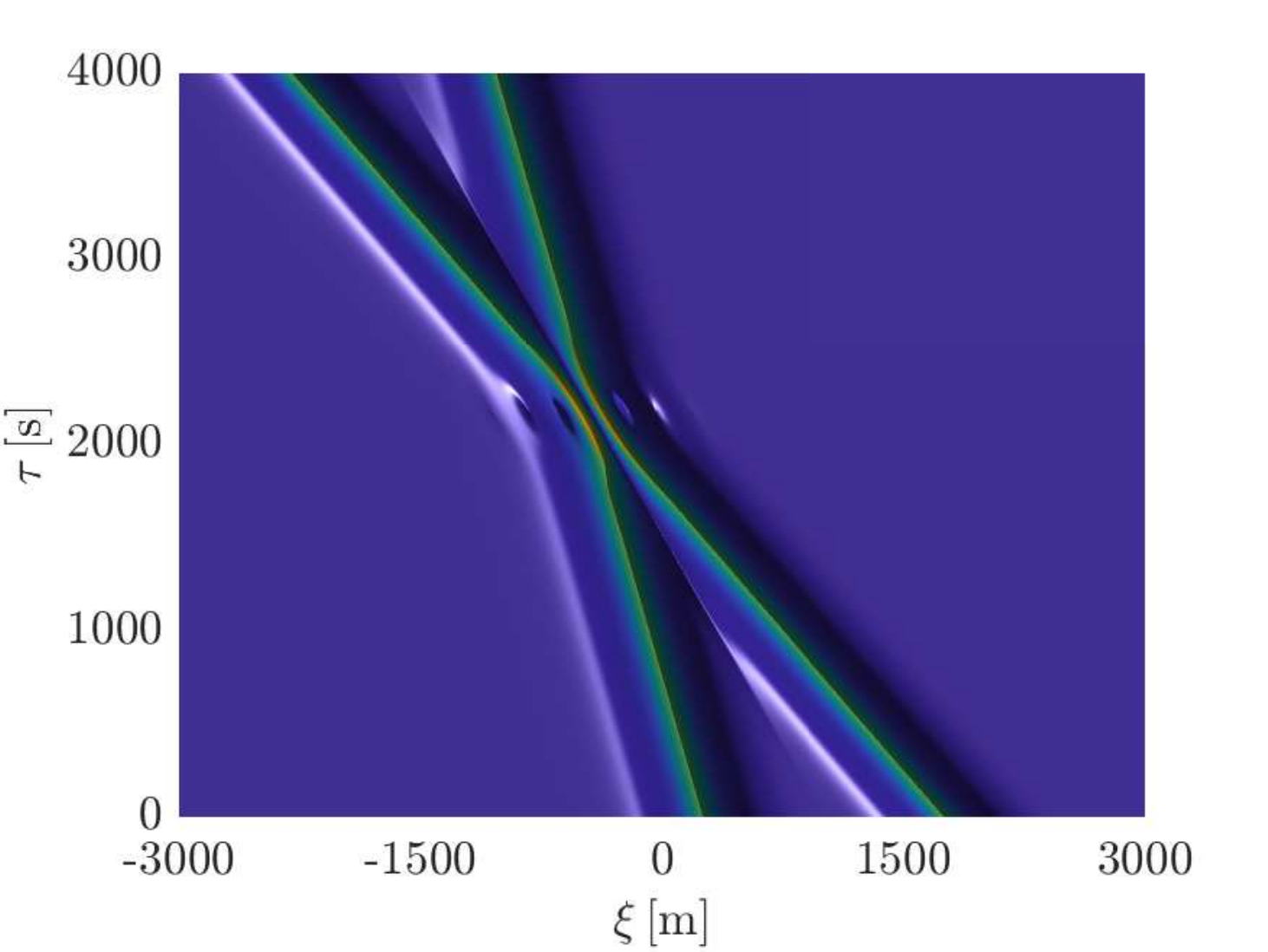}}
	\caption{Interaction behavior of solitons with amplitudes $a_0^1=a_0^2=1 \,\textup{m}$ and velocities $v^1=3v^2= -1 \,\textup{m/s}$ with resulting ${\text{M} =1.5717 \,\textup{m}}$.}
	\label{fig:collision_normal_velo_diff}
\end{figure}

Different soliton interaction scenarios are presented in Figs. \ref{fig:collision_normal}, \ref{fig:collision_normal_ampli_diff} and \ref{fig:collision_normal_velo_diff}, where each solution is shown from two different perspectives. Figure \ref{fig:collision_normal} shows an interaction scenario of solitons with equal wave envelope amplitude and equal velocity but with different directions of movement. In Fig. \ref{fig:collision_normal_ampli_diff}, however, the amplitudes are adjusted such that a scenario with different wave amplitudes is considered. In contrast to the previous figures, Fig. \ref{fig:collision_normal_velo_diff} provides an insight into an interaction scenario of solitons with the same amplitude and the same direction of movement, but with different velocities.
In each scenario considered, a significant rise in the amplitude of the wave envelope can be observed around the time of collision, but its development and magnitude depends on the chosen initial condition. As a result, the presented simulations have indicated that the amplitudes and the directions of movement of the solitons have a considerable influence on the interaction behavior. Before and after the collision, however, the solitons show an unchanged course. 

\begin{figure}[H]
	\centering
	\subfigure{\includegraphics[width=0.49\textwidth]{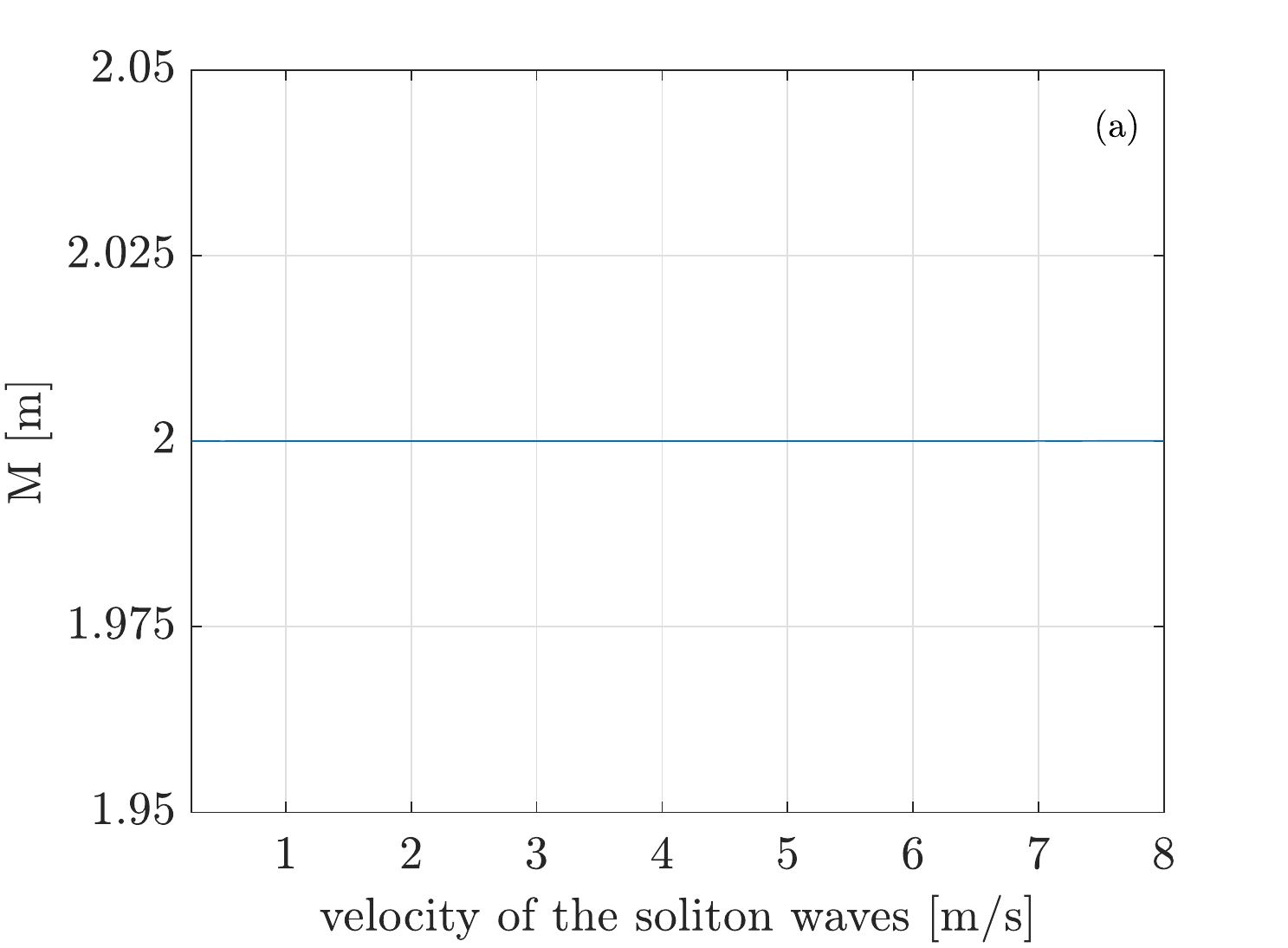}}\label{fig:check_diff_velo_dev_a}
	\subfigure{\includegraphics[width=0.49\textwidth]{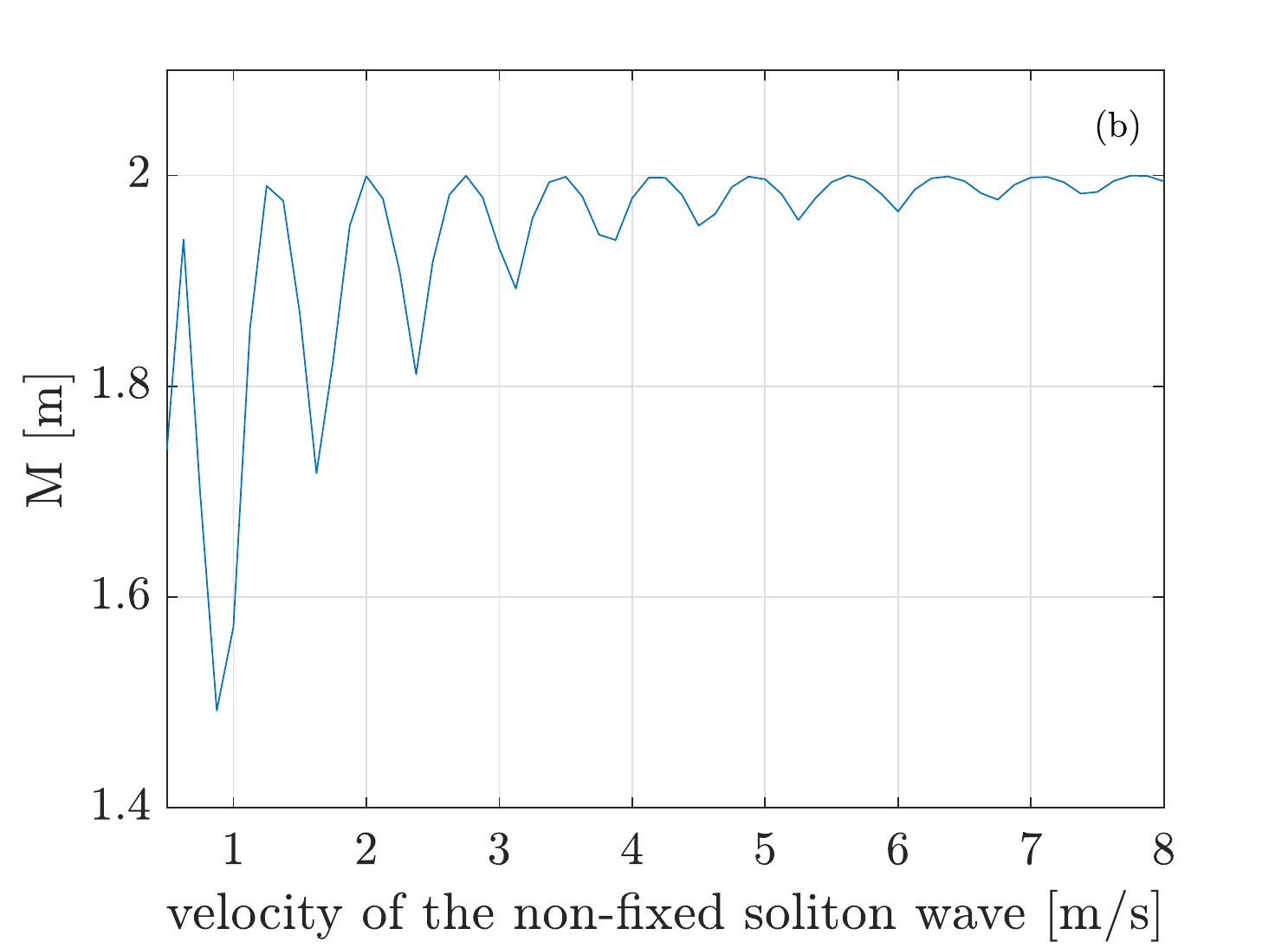}}\label{fig:check_diff_velo_dev_b}
	\caption{Maximal wave envelope amplitude during soliton collision in relation to magnitude of the velocity. Figure \ref{fig:check_diff_velo_dev}a presents the results of solitons with opposite and Fig. \ref{fig:check_diff_velo_dev}b with the same direction of movement.}
	\label{fig:check_diff_velo_dev}
\end{figure}

Additional simulations presented in Fig. \ref{fig:check_diff_velo_dev} have revealed that when a scenario of two waves moving towards each other is considered, the magnitudes of the velocities are a negligible factor with respect to the evolution of the maximal wave envelope amplitude. However, if the solitons have the same direction of motion, a change in the magnitude of the velocities will result in different peaks of the wave envelope amplitude at the time of collision. In general, the peaks in these scenarios are limited by the peak of the envelope amplitude of the scenario presented in Fig. \ref{fig:collision_normal}.

While computing the result in Fig. \ref{fig:check_diff_velo_dev}a, both magnitudes of the velocities are varied equally. In contrast to this, Fig. \ref{fig:check_diff_velo_dev}b was generated by fixing the velocity of the inner soliton to $v^2= -\frac{1}{3} \,\textup{m/s}$, whereas the velocity $v^1$ of the outer wave was varied.

With regard to the topic of extreme waves, it can be assumed that scenarios of the kind shown in Fig. \ref{fig:collision_normal} offer the greatest potential for the creation of extreme waves.
Scenarios of the type shown in Fig. \ref{fig:collision_normal_velo_diff} can also produce comparable peaks of the wave envelope amplitude. In this case, however, the development of the peaks depends very much on the appropriate choice of the initial condition, such that the scenarios mentioned in Fig. \ref{fig:collision_normal} can be assumed to be more suitable.
\subsection{Interaction of disturbed solitons}\label{sec:irregular_inter}
The aim of this section is to make general statements about the interaction behavior in an irregular or random sea state. For this purpose the initial condition of the regular case described in Eq. \eqref{eq:initial_cond_super} is modified according to Eq.~\eqref{eq:initial_condition_irregular}. 
The result of this procedure is illustrated in Fig. \ref{fig:perturbed_IC_2sol}.

\begin{figure}[H]
	\centering
	\subfigure{\includegraphics[width=0.54\textwidth]{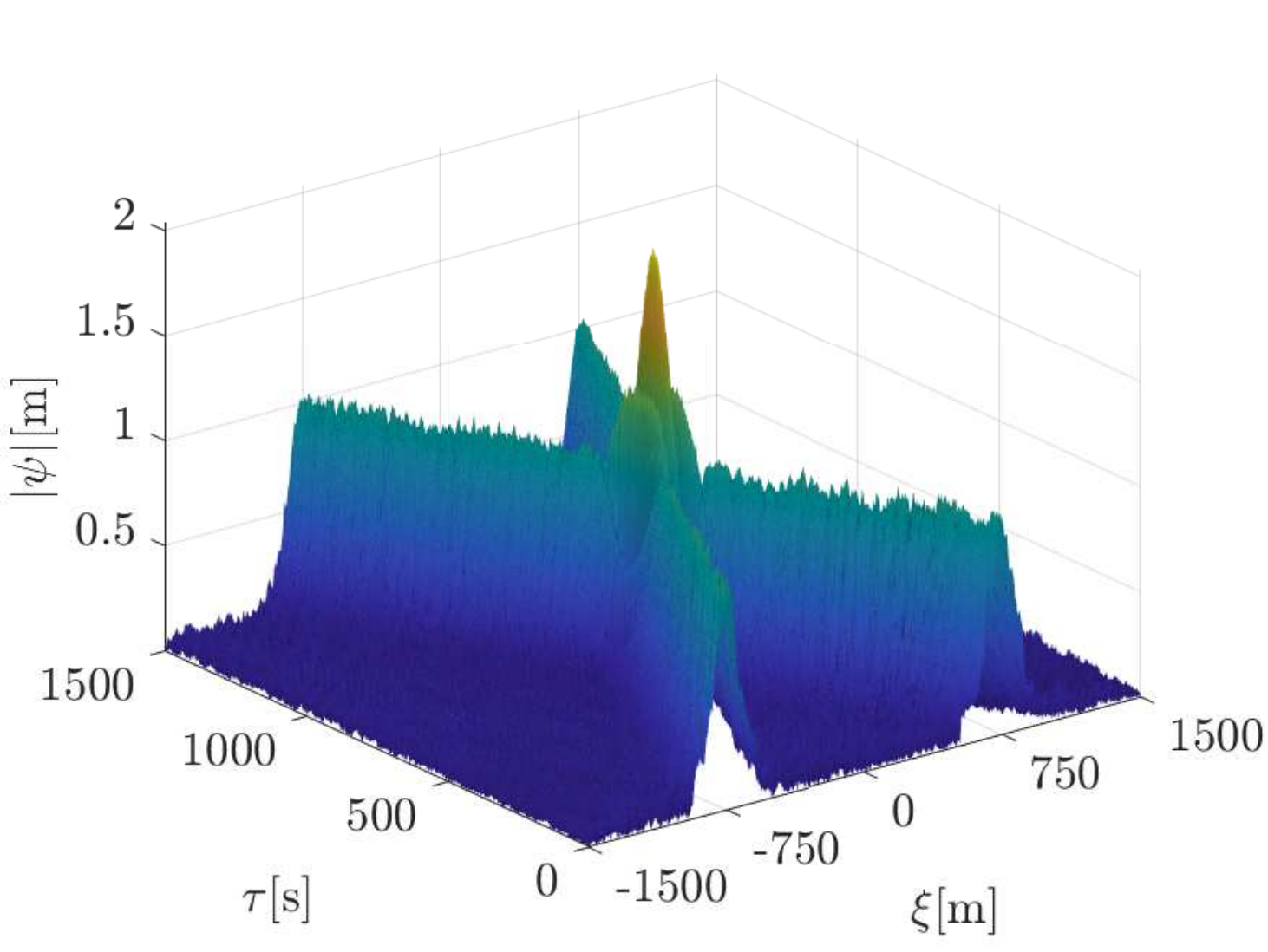}}
	\subfigure{\includegraphics[width=0.45\textwidth]{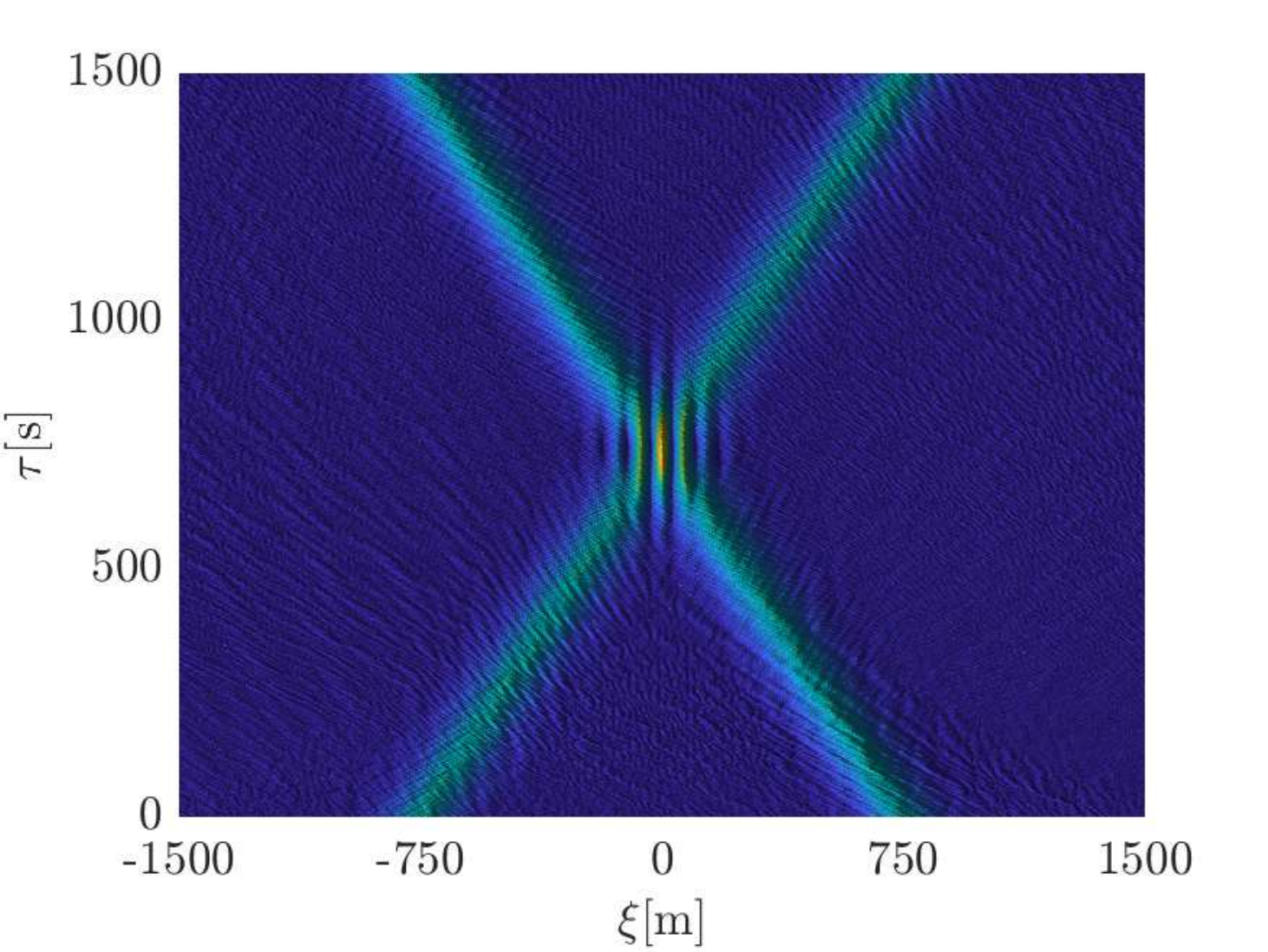}}
	\caption{Interaction behavior of disturbed solitons with amplitudes $a_0^1=a_0^2=1 \,\textup{m}$ and velocities $v^1=-v^2= -1 \,\textup{m/s}$ with resulting ${\text{M} =2.0364 \,\textup{m}}$. The corresponding solution without an initial disturbance can be seen in Fig. \ref{fig:collision_normal}.}
	\label{fig:evo_sol_1}\end{figure}

The development of disturbed solitons shown in Fig. \ref{fig:evo_sol_1} illustrates the fundamental difference between the development of solitons in irregular and regular waves. The constant wave envelope amplitude outside the collision period, which is characteristic for the soliton, is replaced by an oscillating behavior in the irregular case. Identical parameters for the initial condition were chosen for all simulations. This implies that the differences in oscillatory behavior are generated by the stochastic perturbation according to Eq. (\ref{eq:irregular_psi}). This justifies the importance of an empirical analysis.

\begin{figure}[H]
	\centering
	\subfigure{\includegraphics[width=0.39\textwidth]{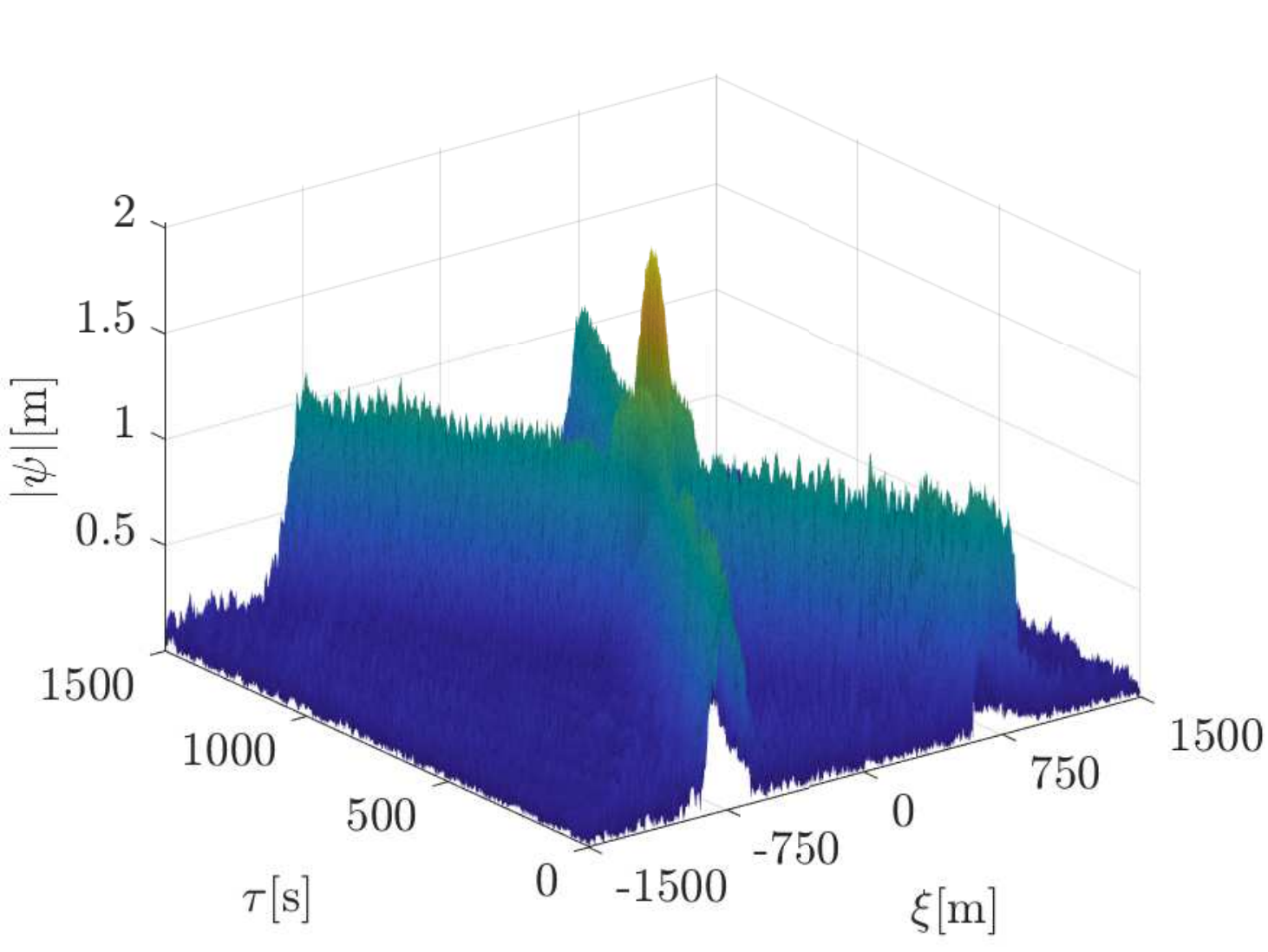}}
	\subfigure{\includegraphics[width=0.39\textwidth]{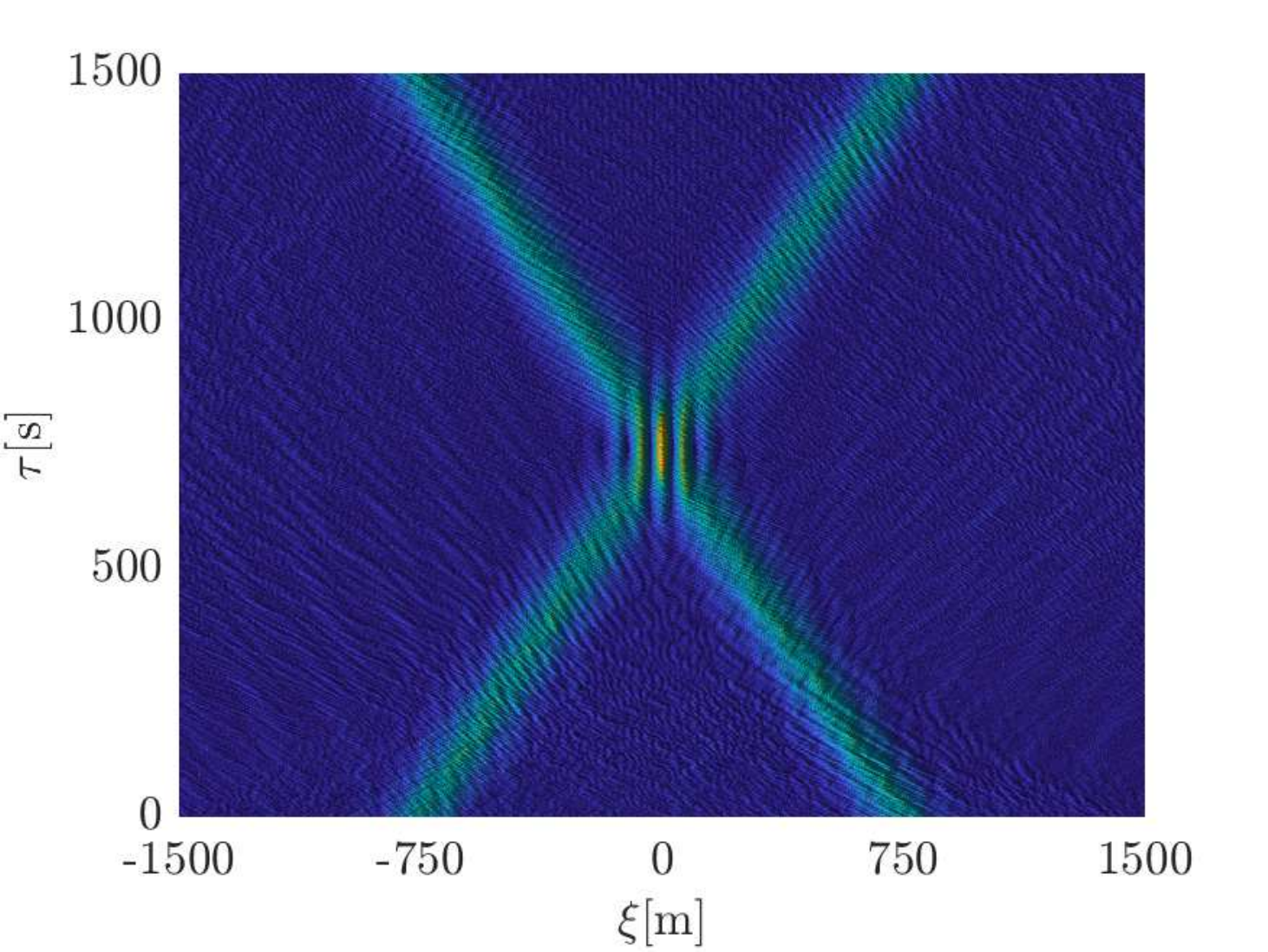}}
	\subfigure{\includegraphics[width=0.39\textwidth]{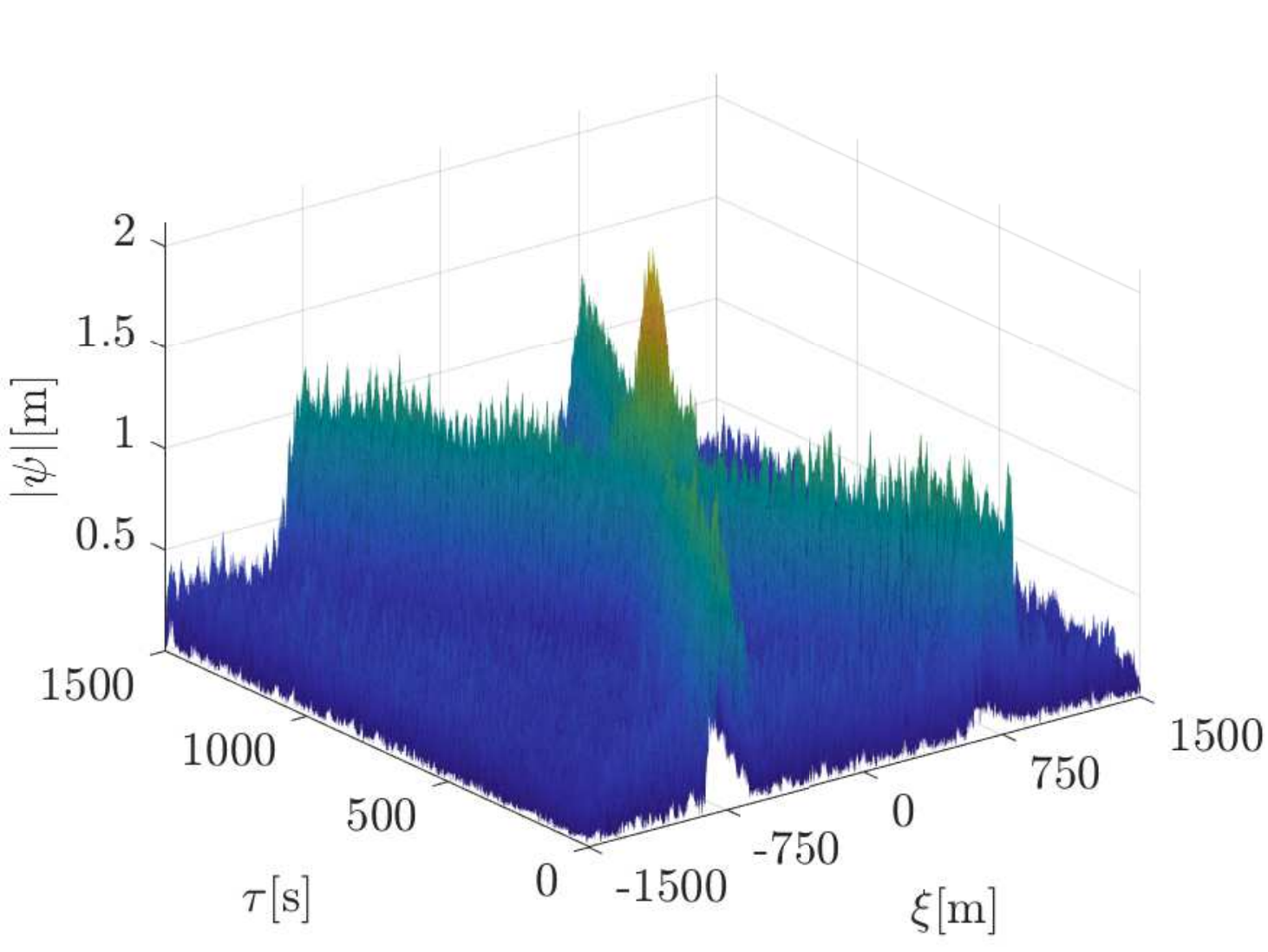}}
	\subfigure{\includegraphics[width=0.39\textwidth]{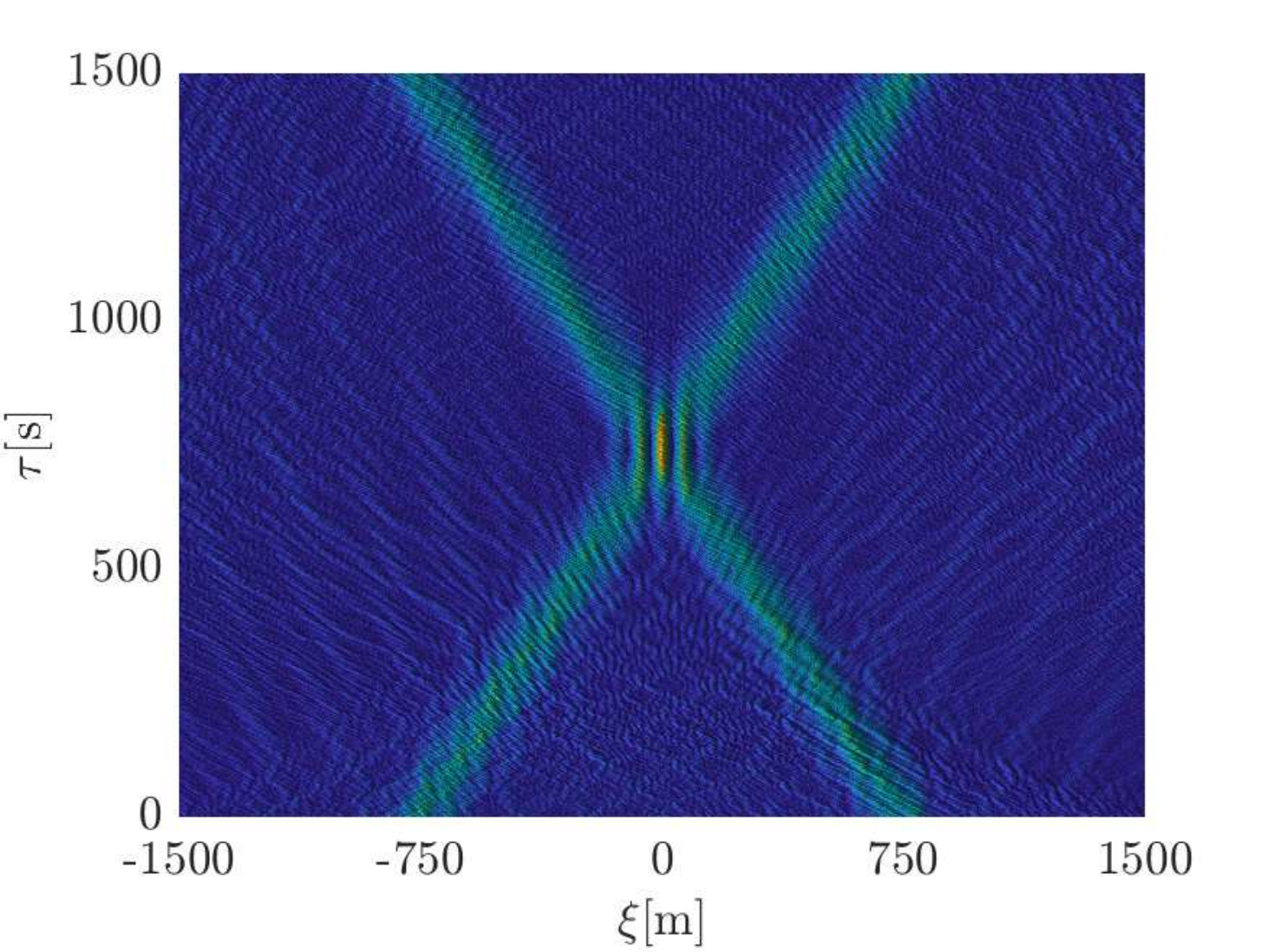}}
	\subfigure{\includegraphics[width=0.39\textwidth]{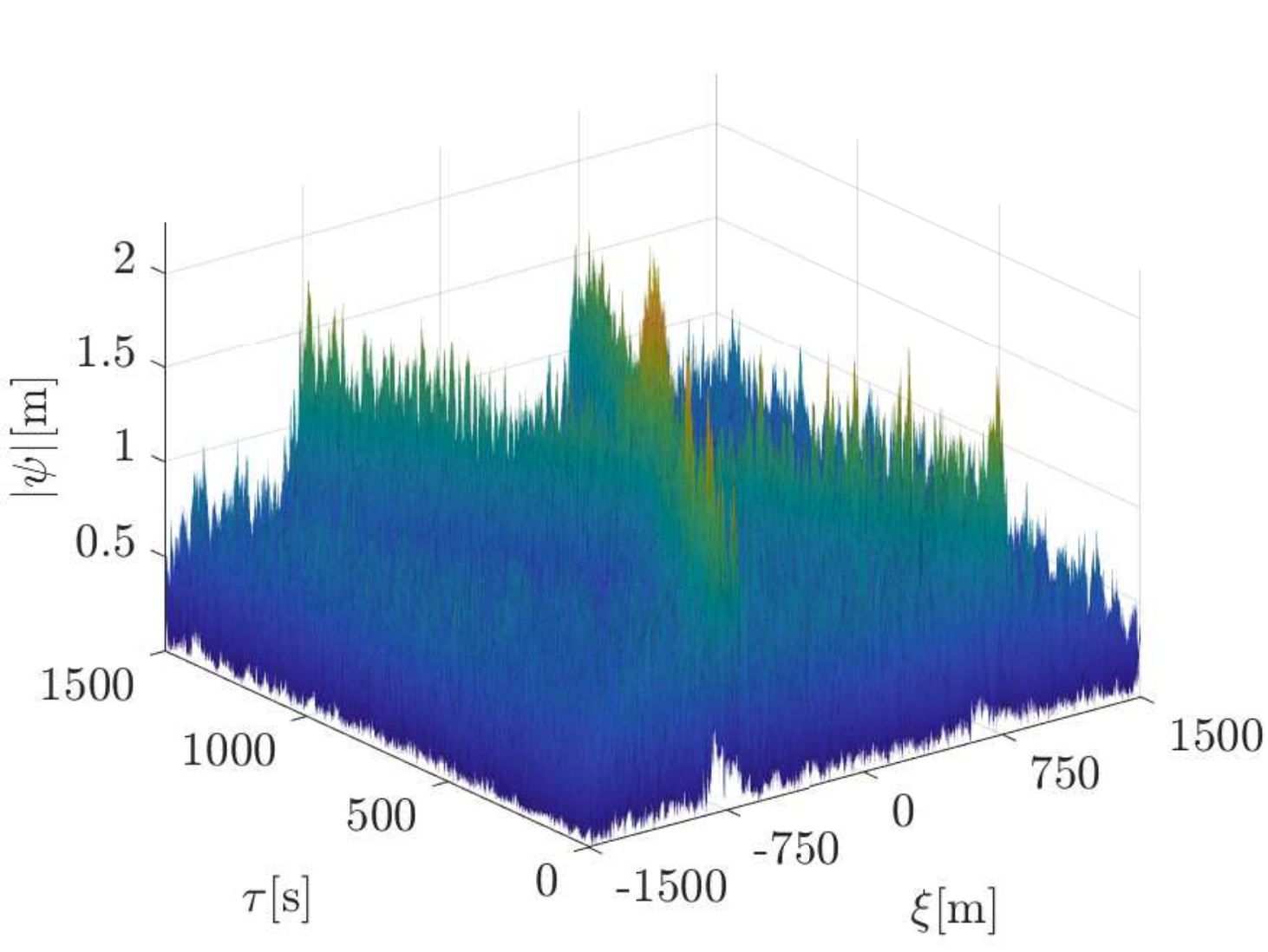}}
	\subfigure{\includegraphics[width=0.39\textwidth]{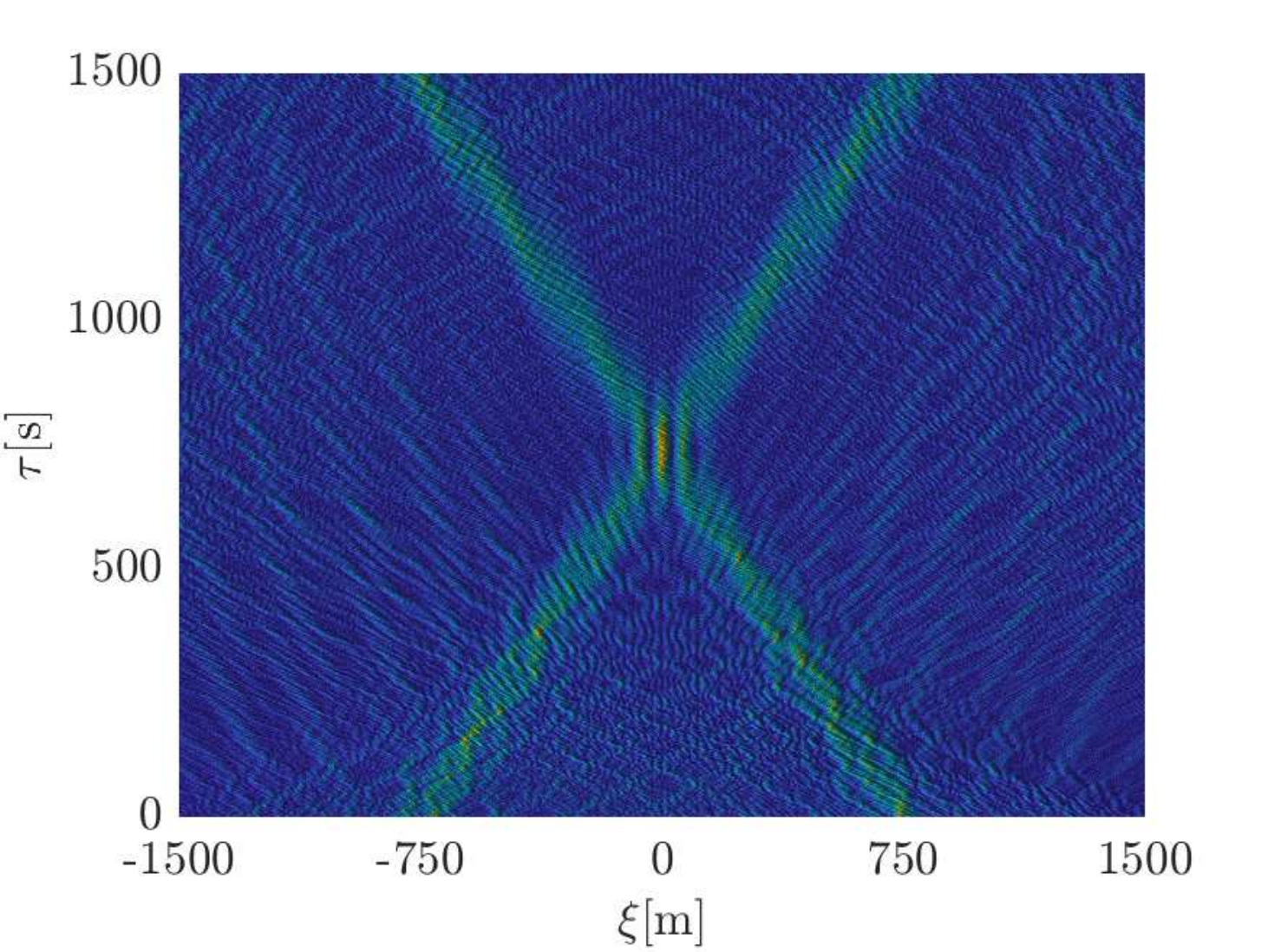}}
	\subfigure{\includegraphics[width=0.39\textwidth]{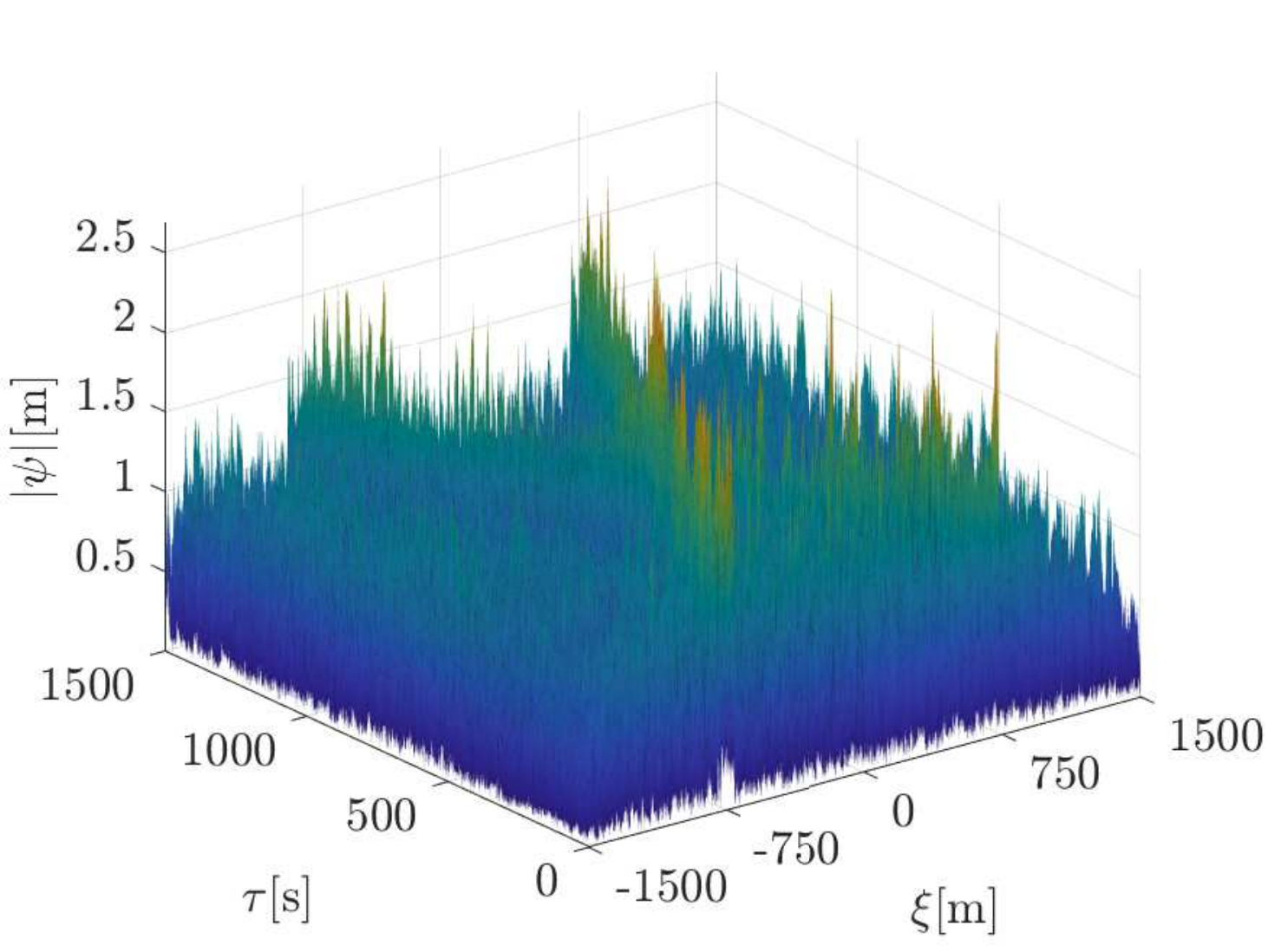}}
	\subfigure{\includegraphics[width=0.39\textwidth]{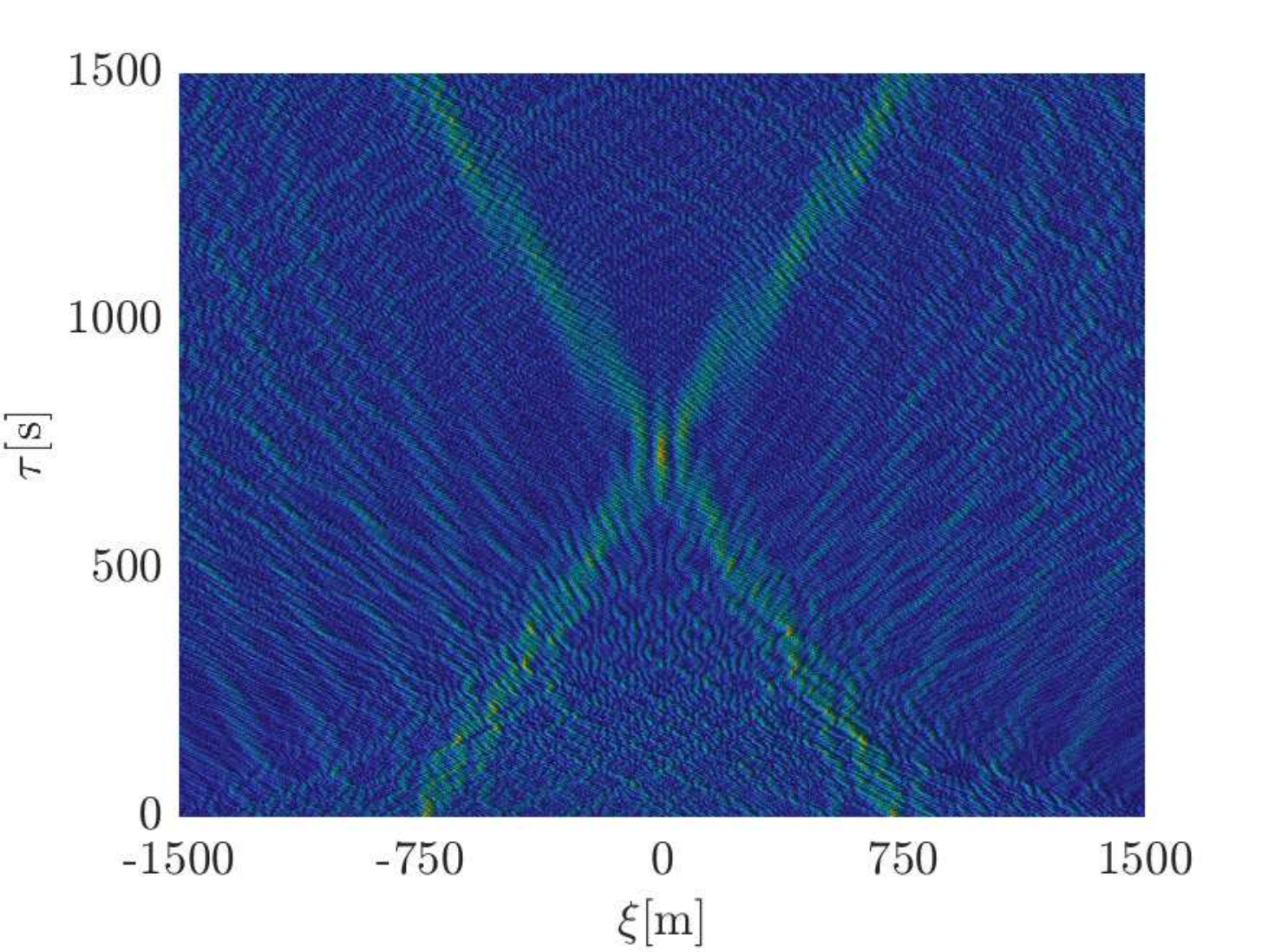}}
	\caption{Interaction behavior of disturbed solitons with amplitudes $a_0^1=a_0^2=1 \,\textup{m}$, velocities $v^1=-v^2= -1 \,\textup{m/s}$ and varying significant wave heights $H_s^1=0.4 \,\textup{m}$, $H_s^2=0.8 \,\textup{m}$, ${H_s^3=1.6 \,\textup{m}}$ and $H_s^4=2.4 \,\textup{m}$. The corresponding maxima of the wave envelope amplitude are given by ${\text{M}^1 = 2.0243 \,\textup{m}}$, ${\text{M}^2 = 2.1240 \,\textup{m}}$, ${\text{M}^3 = 2.2861 \,\textup{m}}$ and ${\text{M}^4 = 2.7181 \,\textup{m}}$.}
	\label{fig:evo_sol_Hs}
\end{figure}

\begin{figure}[H]
	\centering
	\includegraphics[width=1\textwidth]{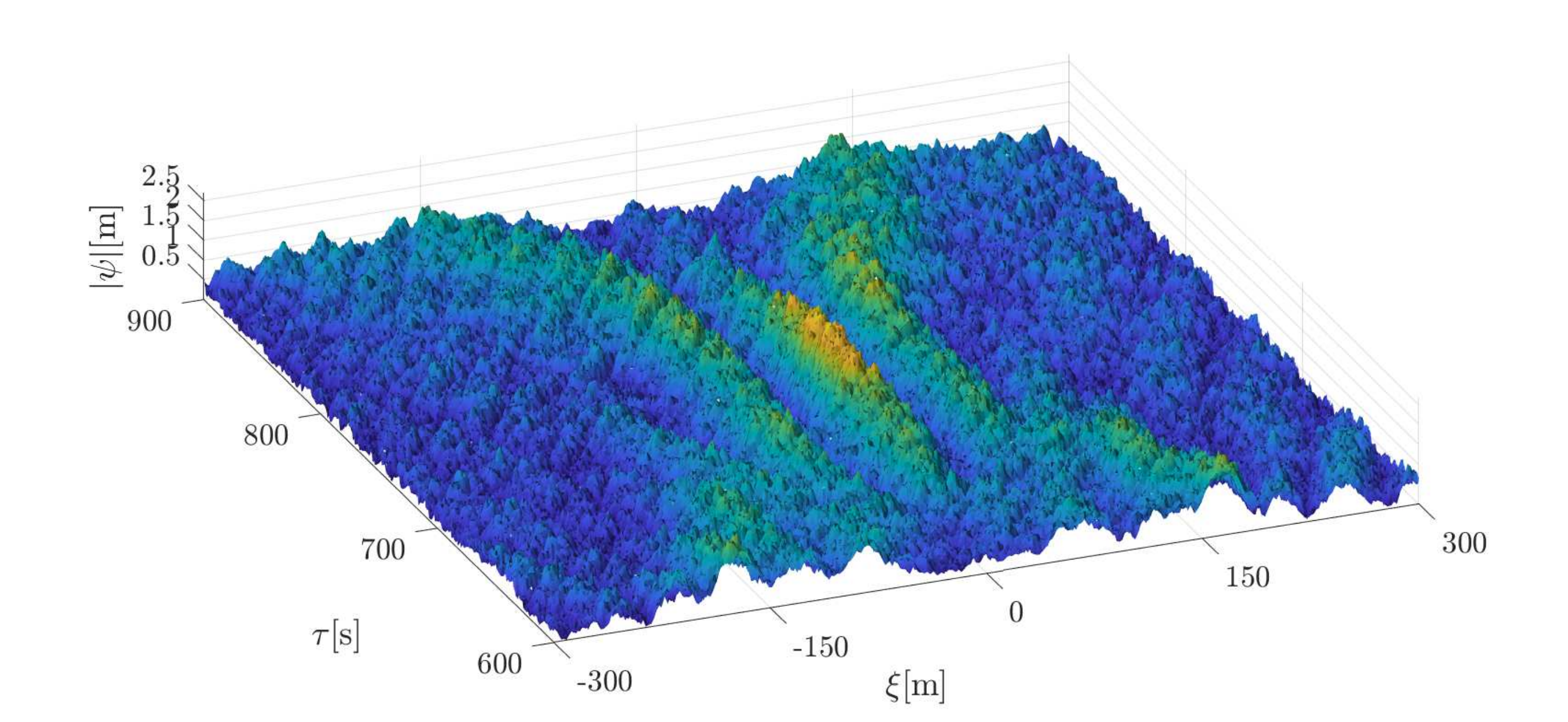}
	\caption{Zoom into the result for $H_s^4=2.4 \,\textup{m}$ depicted in Fig. \ref{fig:evo_sol_Hs}.}
	\label{fig:evo_sol_Hs_zoomed}
\end{figure}

\textcolor{black}{Furthermore, the impact of disturbances with varying significant wave heights on the soliton course is examined in Fig. \ref{fig:evo_sol_Hs} and Fig. \ref{fig:evo_sol_Hs_zoomed}. The major focus is on the effects caused by an increase of the significant wave height. Although no empirical analysis has been performed in this context, the results indicate an enhancement of the maximum of the wave envelope amplitude with rising significant wave height. Moreover, a minor loss of the general shape of the soliton wave is observed in this scenario.}

In order to carry out the analysis of the irregular case, the results of the regular case are considered first. Here it was verified that the scenario of interaction of two solitons with the same amplitude and same velocity but with opposite direction of movement offers the greatest potential for the development of greater wave envelope amplitudes. This information leads the focus on this scenario in the following analysis.

\textcolor{black}{For further reduction of the analysis}, it is first shown that, despite the oscillating wave behavior, the time that elapses before the interaction has a negligible influence on the actual interaction and the resulting wave envelope elevation.

\begin{figure}[H]
	\centering
	\includegraphics[width=0.75\textwidth]{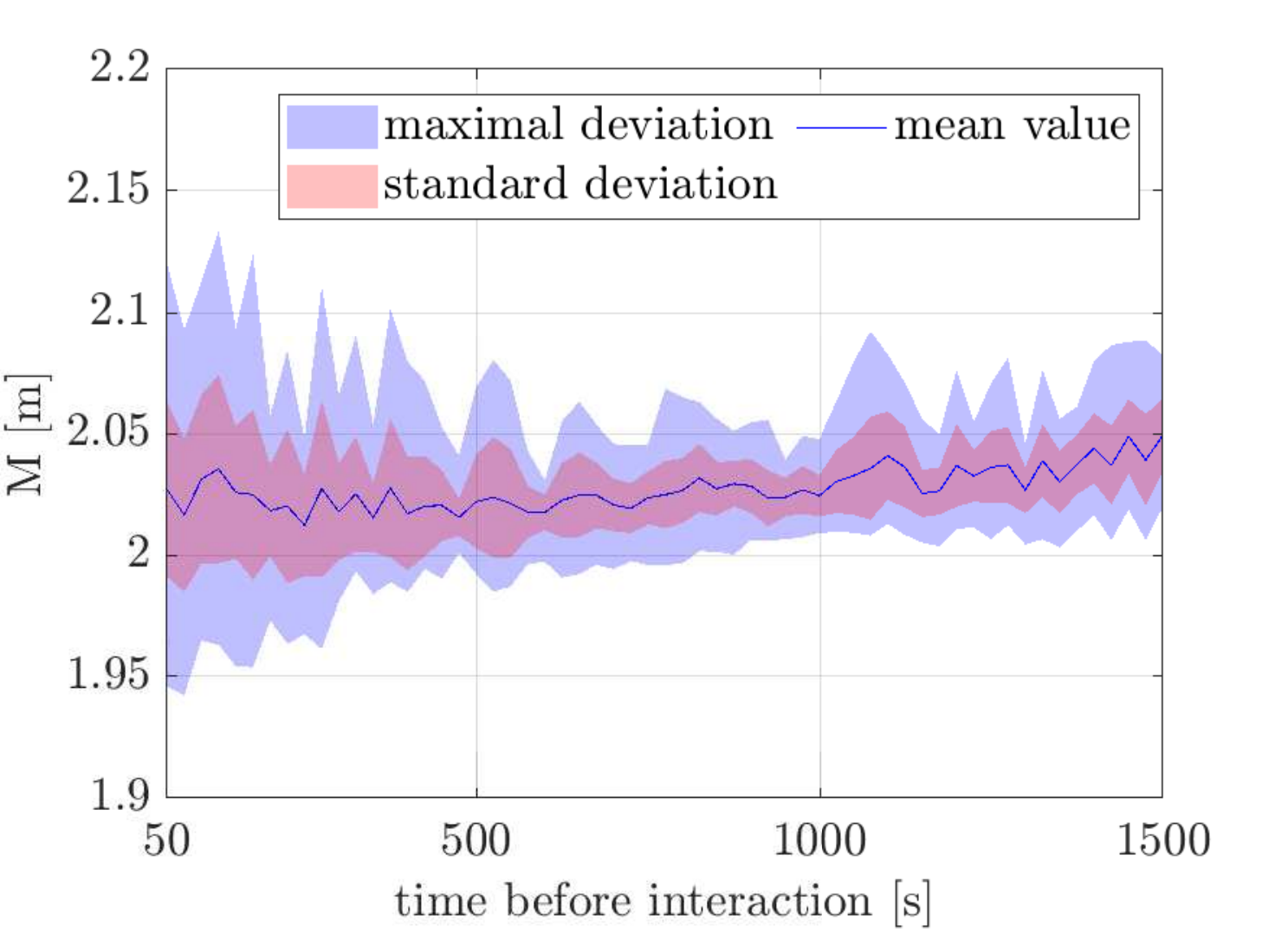}
	\caption{Maximal wave envelope amplitude during interaction for varying time passing before the interaction. For each time \textcolor{black}{100} simulations were performed.}
	\label{fig:time_independecy}
\end{figure}

Since an irregular sea state is considered here, this is shown empirically in Fig. \ref{fig:time_independecy}, which presents the maximum wave envelope amplitude during the interaction as a function of the time elapsing before the interaction. The figure indicates that the average wave elevation has an approximately constant value and furthermore that the standard deviation and the extrema of the maximum wave envelope amplitude are bounded nearly equally for all times. This leads to the conclusion that the time elapsing before the interaction is negligible with respect to the expected wave envelope amplitude.  

However, the visualization in Fig. \ref{fig:time_independecy} only allows statements to be made about the average and limits of the wave envelope amplitude in irregular sea states,  but no statements can be deduced about a stochastic distribution of these. 

\begin{figure}[H]
	\centering
	\includegraphics[width=0.75\textwidth]{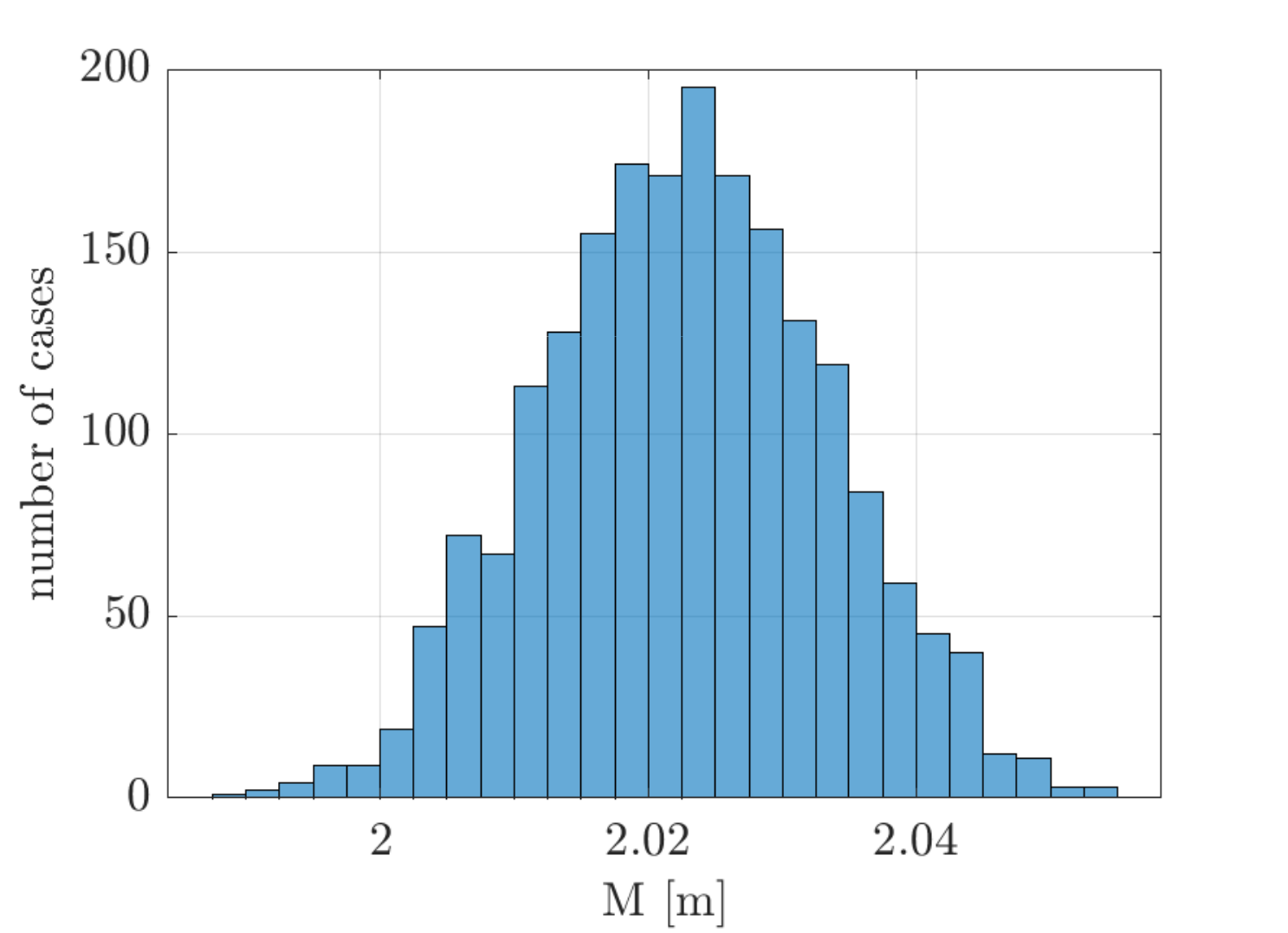}
	\caption{Histogram of the maximal wave envelope amplitude during interaction for a total of 2000 simulations.}
	\label{fig:histogram}
\end{figure}

\begin{figure}[H]
	\centering
	\subfigure{\includegraphics[width=0.49\textwidth]{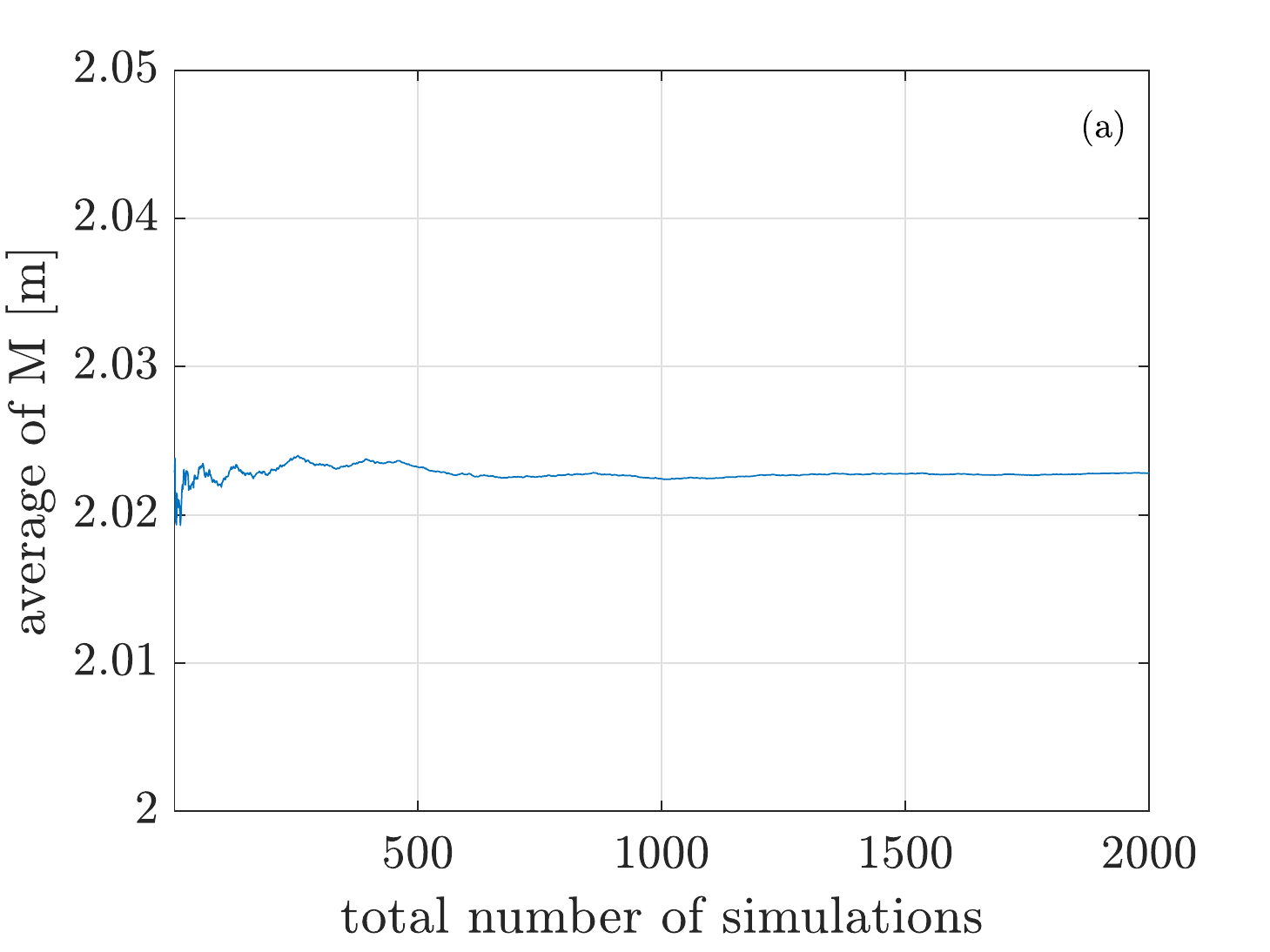}}\label{fig:irreg_evo_a}
	\subfigure{\includegraphics[width=0.49\textwidth]{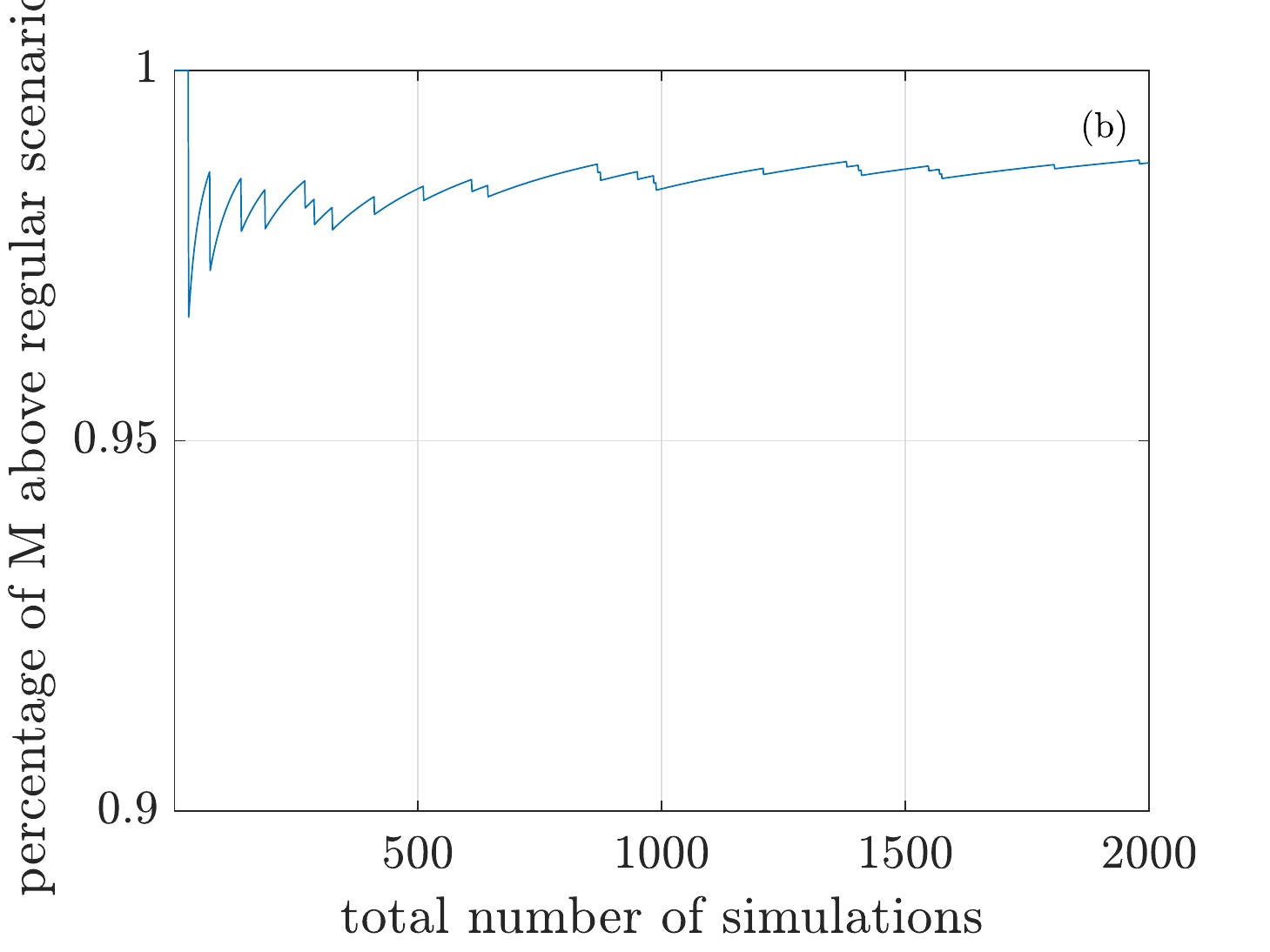}}\label{fig:irreg_evo_b}
	\caption{(a) Average maximal wave envelope amplitude and (b) percentage with maximum wave envelope amplitude above the amplitude of the regular case with respect to the number of simulations.}
	\label{fig:irreg_evo}
\end{figure}

Nevertheless, due to the previous statements, this empirical analysis can be reduced to only one interaction scenario. For this purpose, the scenario presented in Fig. \ref{fig:evo_sol_1} is selected. The main focus is again on the maximum wave envelope amplitude. Therefore, the histogram shown in Fig. \ref{fig:histogram} represents the absolute elevation of the maximal wave envelope amplitude that occurred in the interaction. Thus, this provides insight into the corresponding stochastic distribution. Together with the results from Fig. \ref{fig:irreg_evo}, which verifies the number of simulations as sufficient, the validity of the histogram is proven. Furthermore, the distribution of the maximal wave envelope can be identified as a normal distribution. It should be emphasized here that the mean value approximately corresponds to the maximal amplitude of the regular interaction. In addition, Fig. \ref{fig:irreg_evo}b reveals that an increased amplitude occurs in over half of all simulations. 


Besides the analysis of the interaction itself, a further point of investigation is the impact of the interaction on the disturbed solitons. At first the degree to which the interaction affects the movement of the disturbed soliton is studied. 

This is accomplished by comparing the positions of two disturbed solitons at the end of the simulation period, whereby one of the waves has undergone an interaction and the other one has taken an isolated course. From any other point of view, the waves are chosen identically. The wave position is characterized by the location of the maximum wave envelope amplitude of the disturbed soliton. In the following, the scenario shown in Fig. \ref{fig:evo_sol_1} is used, where in the interaction case the disturbed solitons with negative velocity is analyzed. For the isolated case the other wave with positive velocity is eliminated. 

\begin{figure}[H]
	\centering
	\subfigure{\includegraphics[width=0.49\textwidth]{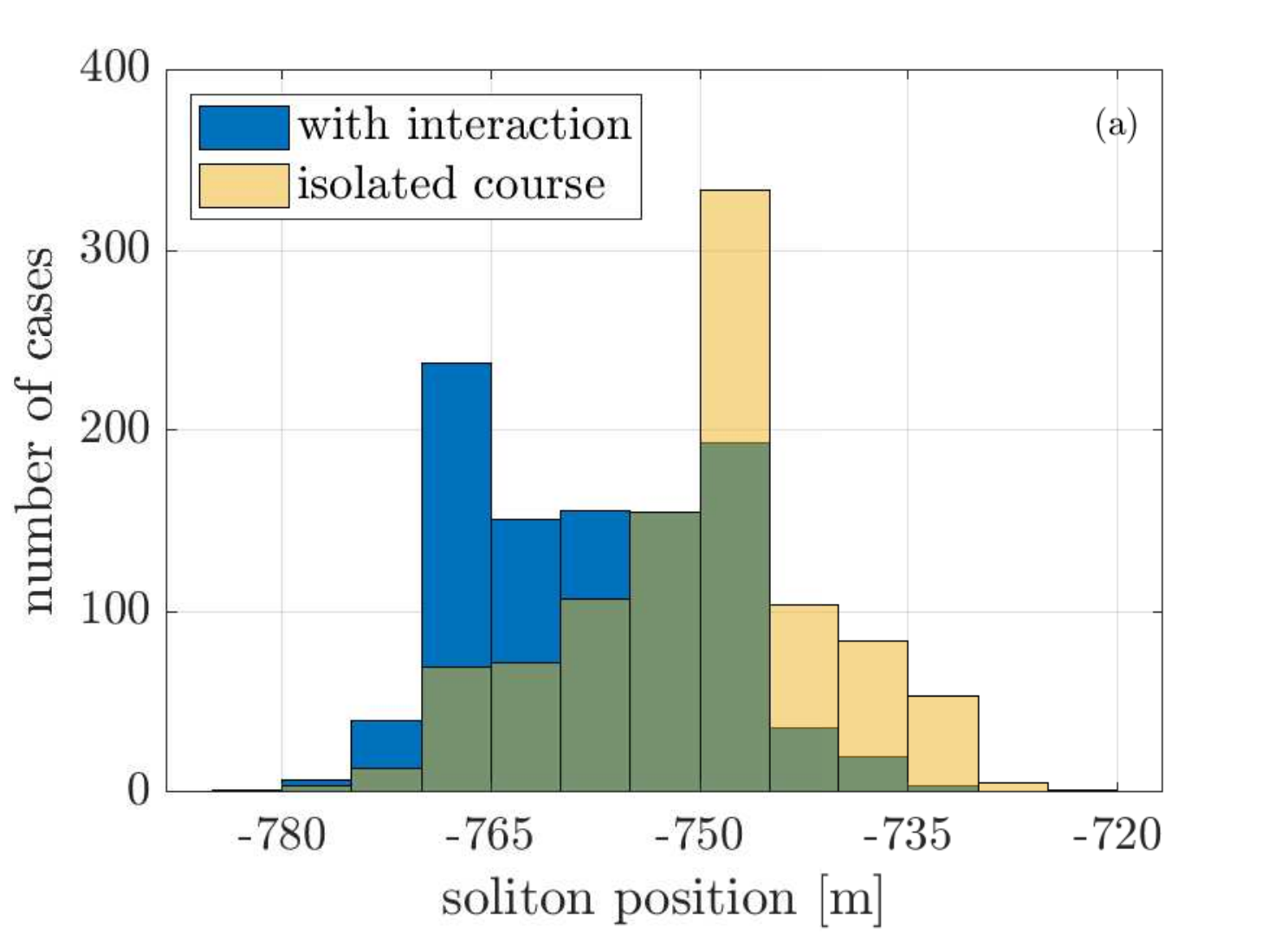}}
	\subfigure{\includegraphics[width=0.49\textwidth]{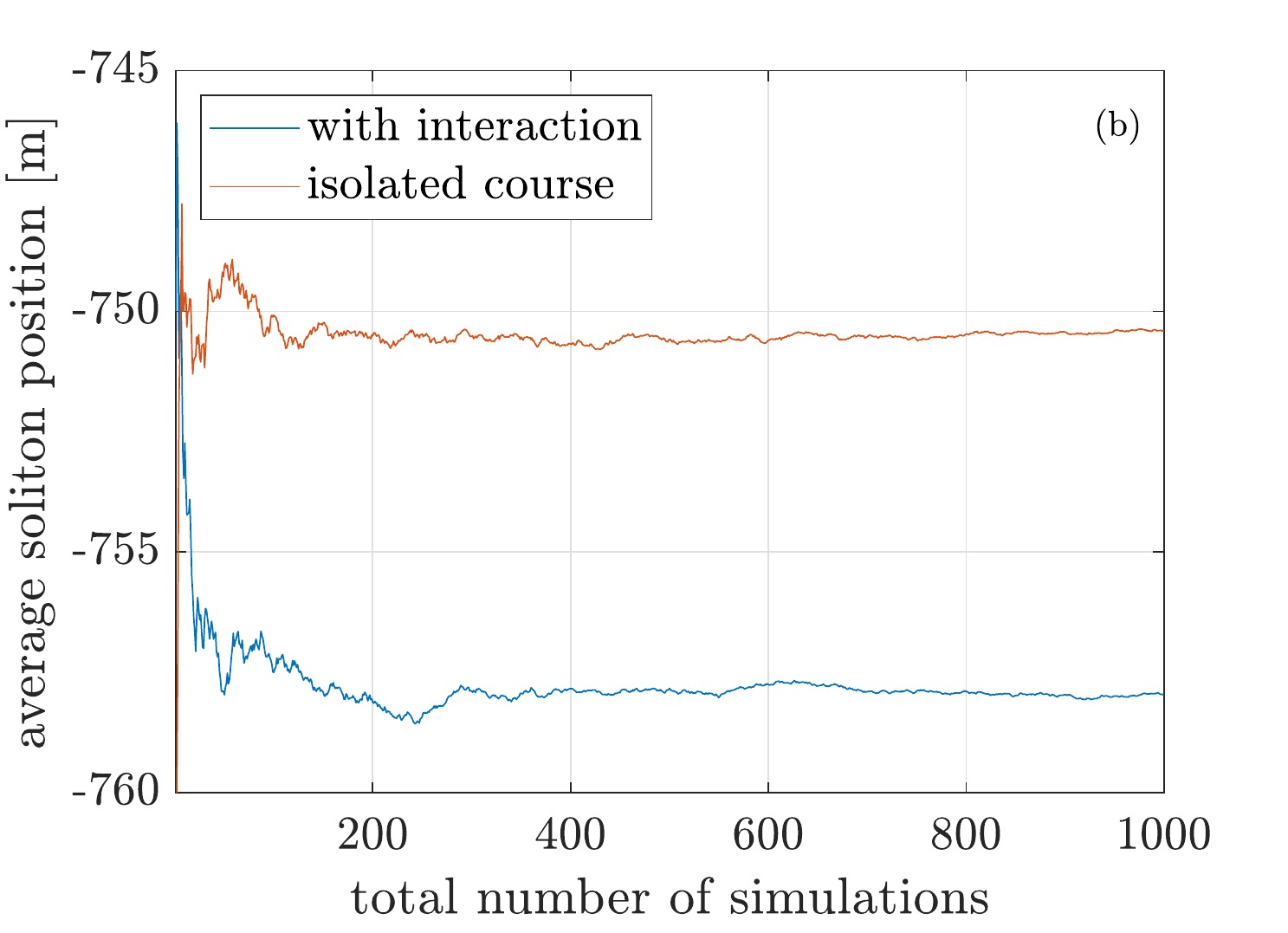}}
	\caption{(a) Histogram of disturbed soliton positions for the isolated and interaction exposed case and (b) the average disturbed soliton position depending on the number of simulations, for a total of 1000 simulations.}
	\label{fig:impact_inter}
\end{figure}

In the regular case in Fig. \ref{fig:collision_normal}, no interaction effect can be determined and due to the symmetric course, the final position is $\xi = 750 \,\textup{m}$. In contrast, the empirical results for the irregular case presented in Fig. \ref{fig:impact_inter} indicate that the interaction is most likely to cause a spatial shift in the direction of movement of the disturbed soliton. However, opposite shifts are also possible, but rather unlikely. Moreover, the average position of the disturbed solitons in the isolated case is almost identical to the regular case. 

But, according to the scale of the spatial simulation area, all shifts of this magnitude must be interpreted as minor.

Finally, beyond the impact of the interaction on the spatial displacement of the disturbed soliton wave, the impact of the interaction on the amplitude of the wave envelope is investigated. For this purpose, the averaged amplitude of the disturbed soliton before and after the interaction is compared. The average amplitude before the interaction is calculated in the time interval $[0 \,\textup{s}, 500 \,\textup{s}]$ and after the interaction in the time interval $[1000 \,\textup{s}, 1500 \,\textup{s}]$, since these intervals can be regarded as free of interactions. 

\begin{figure}[H]
	\centering
	\subfigure{\includegraphics[width=0.47\textwidth]{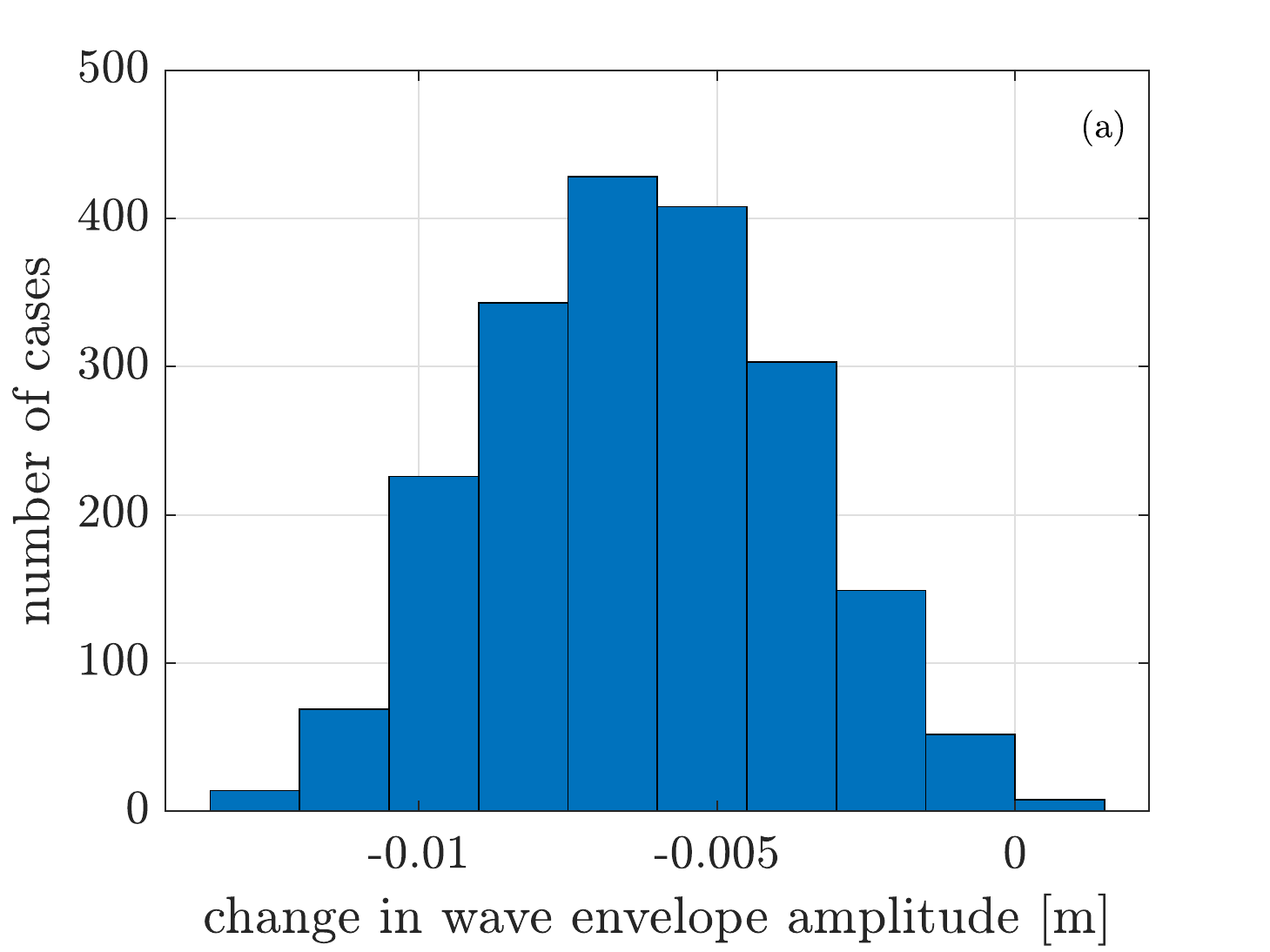}}
	\subfigure{\includegraphics[width=0.51\textwidth]{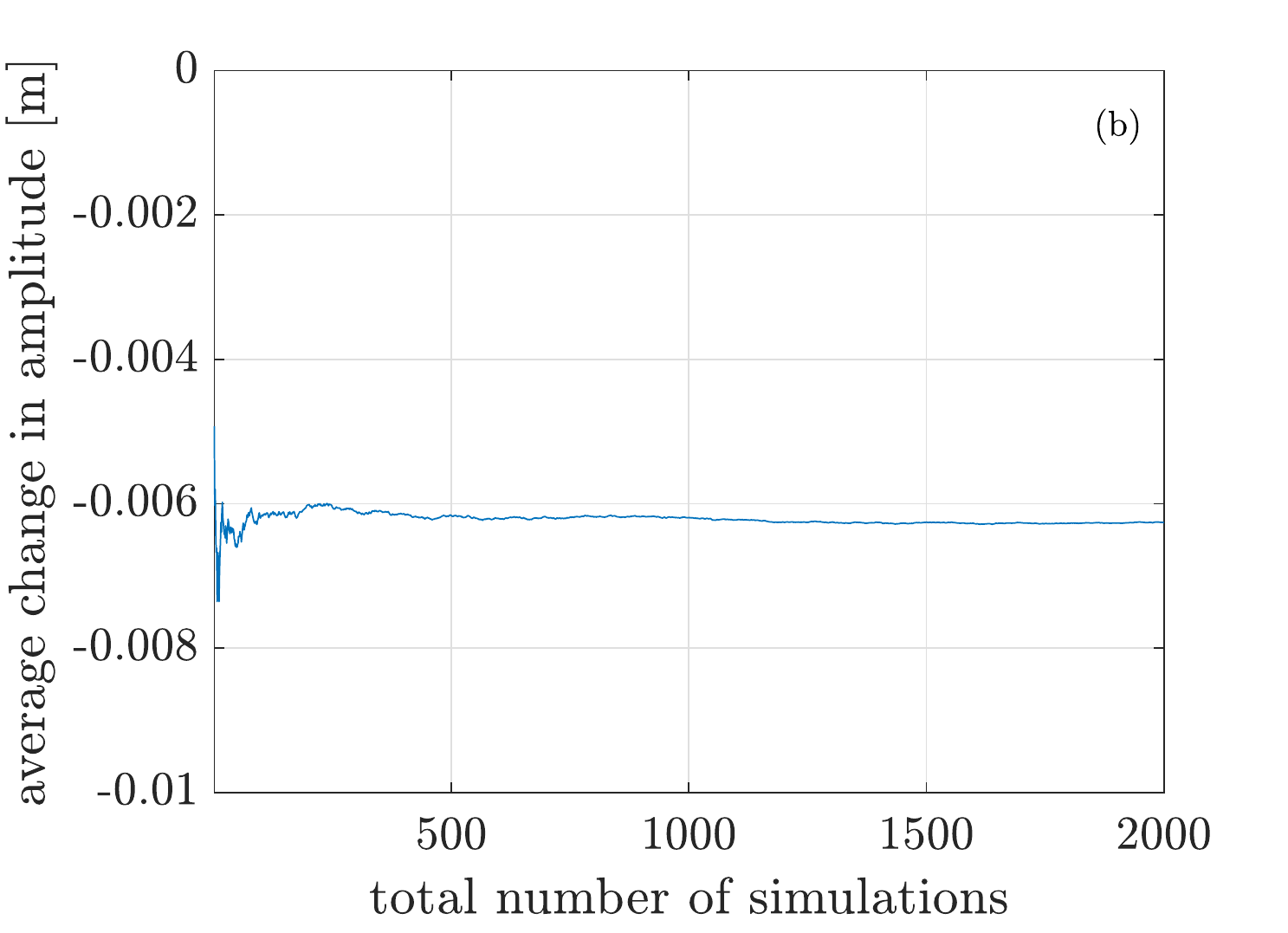}}
	\caption{(a) Histogram of the change in amplitude of the disturbed soliton envelope induced by the interaction and (b) the average amplitude change depending on the number of simulations, for a total of 2000 simulations.}
	\label{fig:impact_inter_height}
\end{figure}

The results summarized in Fig. \ref{fig:impact_inter_height}a reveal that from a statistical perspective and taking the scale of the maximal amplitude into account, the interaction in most cases has only a very slight influence on the disturbed soliton envelope amplitude. On average, Fig. \ref{fig:impact_inter_height}b identifies the influence of the interaction by a slight decrease in amplitude.

\section{Conclusions}\label{sec:Conclusions}
The analysis of the soliton collision in a regular sea state confirms the scenario of solitons moving towards each other at the same speed and with the same amplitude as the scenario with the best potential for the development of an extreme wave. Thus, the empirical analysis of the collision of disturbed solitons in a random sea is based on exactly this scenario. For this, we presented an approach which allows a regular sea to be transformed into a random sea.

Overall, it can be summarized that the amount of time before the disturbed solitons collide in a random sea makes a negligible contribution to the development of an extreme wave. Also the examined impact of the collision on changes in the amplitude and spatial displacement of the disturbed soliton  can be considered insignificant. 

Nevertheless, the application of a realistic random sea state enhances the development of larger wave envelope amplitudes during the soliton collision compared to the regular sea state. Although the maximal amplitude for the regular and irregular state are almost identical on average, the distribution in the irregular case can be approximately identified as a normal distribution.

Applied to the topic of extreme waves, this suggests that a more realistic random sea state can promote the development of these waves and the manifestation of their high amplitude.

\section*{Acknoledgements}
We thank F. Fedele for fruitful discussions about soliton propagation.

\end{document}